\documentclass[sigconf,9pt]{acmart}

\usepackage{algorithm}
\usepackage{algpseudocode}
\usepackage{framed}
\usepackage{cleveref}
\usepackage{subcaption}
\usepackage[abbreviations]{foreign}

\definecolor{takeawaycolor}{RGB}{230, 230, 255}

\usepackage{amsmath}
\usepackage{pifont}
\usepackage{siunitx}
\usepackage{ifthen}
\usepackage{xspace}
\newboolean{showcomments}
\setboolean{showcomments}{true}
\ifthenelse{\boolean{showcomments}}
{ \newcommand{\mynote}[3]{
		\fbox{\bfseries\sffamily\scriptsize#1}
		{\small$\blacktriangleright$\textsf{\emph{\color{#3}{#2}}}$\blacktriangleleft$}}
	\newcommand{\zzz}[1]{{\setlength{\fboxsep}{2pt}\fcolorbox{black}{yellow}{\textsf{\emph{#1}}}}\xspace}}
{ \newcommand{\mynote}[3]{}
	\newcommand{\zzz}[1]{}}

\definecolor{maroon}{rgb}{0.5, 0.0, 0.0}

\usepackage{acronym}
\acrodef{ANN}{Approximate Nearest Neighbor}
\acrodef{ML}{machine learning}
\acrodef{LLM}{large language model}
\acrodef{RAG}{retrieval-augmented generation}
\acrodef{HNSW}{hierarchical navigable small world}
\acrodef{DHT}{distributed hash table}
\acrodef{NLP}{natural language processing}
\acrodef{BGP}{border gateway protocol}
\acrodef{MeSH}{Medical Subject Headings}
\acrodef{LSH}{locality-sensitive hashing}
\acrodef{NNS}{nearest neighbor search}
\newcommand{\sys}{\textsc{ZeroRAG}\xspace}

\newcommand{\semantica}{\textsc{Semantica}\xspace}
\newcommand{\dhtlsh}{\textsc{DHT-LSH}\xspace}

\newcommand{\aolia}{\textsc{AOL-IA}\xspace}
\newcommand{\amazon}{\textsc{Amazon Electronics}\xspace}
\newcommand{\pubmedqa}{\textsc{PubMedQA}\xspace}
\newcommand{\pubmed}{\textsc{PubMed}\xspace}

\algblockdefx[UPON]{Upon}{EndUpon}[1]{\textbf{On receive} #1}{}
\usepackage{physics}
\usepackage{thm-restate}
\usepackage{bbm}

\allowdisplaybreaks

\graphicspath{ {figures/}{graphs/} }
\usepackage{tikz}
\usepackage[eulergreek]{sansmath}
\usepackage{pgfplots}
\usepackage{pgfplotstable}
\usepackage{comment}

\pgfplotsset{compat=newest}
\usepgfplotslibrary{external,units,colorbrewer,groupplots,fillbetween,statistics}
\tikzsetexternalprefix{figures/}
\tikzset{external/mode=list and make}
\usetikzlibrary{patterns,shapes.misc}

\usepackage[eulergreek]{sansmath}
\usepackage{pgfplots}
\usepackage{pgfplotstable}

\pgfplotsset{compat=newest}
\usepgfplotslibrary{units,colorbrewer,groupplots,fillbetween,statistics}
\usetikzlibrary{arrows.meta,backgrounds,calc,fit,shapes.geometric,positioning,patterns,shapes.misc}

\pgfplotsset{
  width=3.25in, height=1.61in,
  every axis/.append style={
    semithick,
    ymajorgrids=true,
    label style={font=\sffamily\small},
    title style={font=\sffamily\small},
  },
  every axis plot/.append style={thick},
  legend style={
    draw=none,
    fill=none,
    text opacity=1,
    anchor=south,
    at={(0.5,1)},
    font=\sffamily\small,
  },
  legend cell align=left,
  legend columns=-1,
  xtick align=outside,
  xtick pos=bottom,
  major tick length=1mm,
  tick label style={font=\sansmath\sffamily\small},
  every axis label={font=\sffamily\small},
  label style={font=\sffamily\small},
  {cycle list/Dark2},
  {cycle list/Set1},
  cycle list name=Dark2,
  cycle multiindex* list={
    Dark2\nextlist
    {solid,dashed,dotted,dashdotted,densely dashdotdotted}\nextlist
    mark list*\nextlist
  },
  select coords between index/.style 2 args={
    x filter/.code={
      \ifnum\coordindex<#1\fi
      \ifnum\coordindex>#2\fi
    }
  },
  discard if not/.style 2 args={
    x filter/.code={
      \edef\tempa{\thisrow{#1}}
      \edef\tempb{#2}
      \ifx\tempa\tempb
      \else
        
      \fi
    }
  },
  sys/.style          ={color=Set1-B, solid,         very thick, no marks},
  sys-dashed/.style   ={color=Set1-B, dashed,        thick,      no marks},
  sys-dotted/.style   ={color=Set1-B, densely dotted,thick,      no marks},
  sys-dashdot/.style  ={color=Set1-B, dashdotted,    thick,      no marks},
  std bars/.style={
    error bars/y dir=both, error bars/y explicit,
    error bars/error bar style={black, thin},
    error bars/error mark options={black, thin, rotate=90, mark size=1.5pt},
  },
  bar legend/.style={
    area legend,
    legend image code/.code={\draw[#1] (0cm,-0.07cm) rectangle (0.16cm,0.07cm);},
  },
}

\def\overleafhome{/tmp}
\newcommand{\inputplot}[1]{%
	\ifx\homepath\overleafhome%
	\includegraphics{#1.pdf}%
	\else%
	{\sffamily\scriptsize
	 \tikzsetnextfilename{#1}%
	 \input{plots/#1/#1}}%
	\fi
}

\newcommand{\newgroupwidth}[2]
{\expandafter\xdef\csname groupwidth#1\endcsname{#2}}

\newcounter{groupwidth}
\newsavebox{\groupwidthbox}
\makeatletter
\newenvironment{groupwidth}[1]
{\edef\groupnumber{#1}
	\stepcounter{groupwidth}%
	\@ifundefined{groupwidth\thegroupwidth}{\pgfmathsetlengthmacro{\mywidth}{\linewidth/\groupnumber}}%
	{\expandafter\let\expandafter\mywidth\csname groupwidth\thegroupwidth\endcsname}%
	\begin{lrbox}{\groupwidthbox}%
		\tikzset{/pgfplots/width={\mywidth}}%
		\ignorespaces}%
	{\end{lrbox}%
	\usebox\groupwidthbox
	\pgfmathsetlengthmacro{\mywidth}{\mywidth + (\linewidth - \wd\groupwidthbox)/\groupnumber}
	\immediate\write\@auxout{\string\newgroupwidth{\thegroupwidth}{\mywidth}}}
\makeatother

\acmConference[Middleware '26]{27th ACM International Middleware Conference}{December
14--18}{Tarragona, Spain}

\copyrightyear{2026}
\acmYear{2026}
\setcopyright{cc}
\setcctype{by}
\acmConference[Middleware '26]{27th International Middleware Conference}{December 14--18, 2026}{Tarragona, Spain}
\acmBooktitle{27th International Middleware Conference (Middleware '26), December 14--18, 2026, Tarragona, Spain}
\acmDOI{10.1145/3801927.3858412}
\acmISBN{979-8-4007-2621-7/2026/12}

\begin{document}

\title[ZeroRAG]{ZeroRAG: Adaptive Semantic Peer Routing for \\Decentralized Retrieval-Augmented Generation}

\author{Petru Neague}
\affiliation{%
  \institution{Delft University of Technology}
  \city{Delft}
  \country{The Netherlands}
}
\email{p.m.neague@tudelft.nl}
\author{Quinten Stokkink}
\affiliation{%
  \institution{Delft University of Technology}
  \city{Delft}
  \country{The Netherlands}
}
\email{q.a.stokkink@tudelft.nl}
\author{Martijn de Vos}
\affiliation{%
  \institution{EPFL}
  \city{Lausanne}
  \country{Switzerland}
}
\email{martijn.devos@epfl.ch}
\author{Rafael Pires}
\affiliation{%
  \institution{EPFL}
  \city{Lausanne}
  \country{Switzerland}
}
\email{rafael.pires@epfl.ch}
\author{Mathis Randl}
\affiliation{%
  \institution{EPFL}
  \city{Lausanne}
  \country{Switzerland}
}
\email{mathis.randl@epfl.ch}
\author{Anne-Marie Kermarrec}
\affiliation{%
  \institution{EPFL}
  \city{Lausanne}
  \country{Switzerland}
}
\email{anne-marie.kermarrec@epfl.ch}
\author{Johan Pouwelse}
\affiliation{%
  \institution{Delft University of Technology}
  \city{Delft}
  \country{The Netherlands}
}
\email{j.a.pouwelse@tudelft.nl}

\renewcommand{\shortauthors}{Neague, et al.}

\begin{abstract}

\Ac{RAG} is a standard approach to enhance the reliability of \ac{LLM} outputs by retrieving documents relevant to a query and supplying them in the \ac{LLM} prompt.
Today's \ac{RAG} systems collect, index, and serve these documents from centralized stores. 
In practice, however, documents are owned by organizations, devices, and individuals who cannot or will not surrender custody of them to a central party, \eg, because of privacy, copyright or regulatory compliance.
Existing modern decentralized retrieval techniques either reintroduce a coordinating authority, or rely on a global data structure that requires continuous network-wide coordination (\eg, DHTs).
We present \sys, a decentralized routing layer for document retrieval in \ac{RAG} that performs semantic query routing across an overlay of peers without global data structures or centralized orchestration.
Documents remain with the peers that own them, and each peer keeps only lightweight local state: a bounded set of neighbor advertisements and a cache of recently routed queries.
Using these, \sys steers each query through a best-first walk that progressively rewires the overlay toward semantically relevant neighbors, so recall improves over time as the network self-organizes.
We evaluate \sys using three real-world workloads and two decentralized baselines.
We find that \sys outperforms baselines in terms of recall and required network messages, approaches the recall and \ac{LLM} answer accuracy of a centralized document store, self-heals after catastrophic peer failures, and scales to \num{100000} peers.

\end{abstract}



\maketitle

\section{Introduction}
\Acfp{LLM} have transformed natural language processing because of their strong capabilities across tasks such as text generation, translation, and summarization~\cite{vaswani2017attention,brown2020language}.
One of the most popular applications of \acp{LLM} is question answering~\cite{zhuang2023toolqa}.
In this domain, a fundamental challenge is to ensure the reliability of \ac{LLM}-generated answers.
However, \Acp{LLM} are prone to hallucinations where they confidently generate false or misleading information~\cite{alansari2026large}.
This makes it difficult to trust the outputs of \acp{LLM} without double-checking factuality against other sources of truth~\cite{lee2024one}.
These reliability concerns are particularly problematic in high-stakes settings, such as healthcare, finance, or law, where inconsistent or fabricated answers can have serious real-world consequences~\cite{ji2023towards}.

\Ac{RAG} is a popular approach to improving \ac{LLM} reliability without retraining or fine-tuning the model~\cite{lewis2020retrieval,fan2024survey}.
Rather than relying solely on parametric knowledge embedded in the model itself, \ac{RAG} first retrieves relevant documents from an external source based on the user's query, then includes them in the prompt before generating a response.
Both queries and documents are typically represented as high-dimensional embedding vectors capturing semantic meaning, and retrieval involves finding the embeddings nearest to the query, a process known as \ac{NNS}~\cite{nene1997simple}.
This allows the \ac{LLM} to ground its answers in up-to-date or domain-specific information, improving factual accuracy and consistency.

Existing \ac{RAG} systems typically assume that documents are collected, indexed, and served from a centralized store under unified administrative control~\cite{lewis2020retrieval,fan2024survey,zhao2026retrieval}.
While this simplifies the \ac{RAG} pipeline and results in low latency for document retrieval, it is not always a viable or desirable architecture.
Data may be inherently distributed across organizations, devices, or individuals who cannot or will not surrender custody of their documents to a central party, whether for reasons of privacy, regulatory compliance, data ownership, or operational boundaries~\cite{RAGRoute,kuo2025research}.
A hospital network, for instance, cannot 
pool patient records into a shared index without additional access control~\cite{hipaa1996,rieke2020future}.
Similarly, a federation of research institutions may each hold proprietary datasets they are unwilling to freely share with other parties.
Beyond ownership concerns, centralizing large, heterogeneous, and continuously evolving document collections introduces infrastructure costs and creates a single point of failure~\cite{gilbert2002brewer,Harmony,spire}.
Despite these limitations, no general-purpose decentralized retrieval infrastructure exists for \ac{RAG}: existing systems either assume centralized control, or distribute retrieval only within tightly managed federated or cloud settings that still require some coordinating authority.



We present \sys, a decentralized routing layer that enables document retrieval by routing queries across a decentralized overlay, without requiring global data structures or centralized orchestration.
Documents remain with the peers that own them and peers themselves are in control whether they want to share documents with the party requesting them.
Retrieval with \sys only uses local coordination.
Each peer in ZeroRAG maintains only lightweight local state: compact summaries of the content exposed by neighboring peers and a bounded cache of previously-routed queries, and uses these to guide incoming queries to peers likely to hold documents relevant to that query. This is a key departure from prior decentralized nearest-neighbor systems, which route based only on what peers \emph{hold}: \sys additionally exploits what peers have recently \emph{asked}, referring query issuers toward peers whose recent query history is semantically aligned, not only their local inventory. For example, REMINDIN'~\cite{Remindin} only learns from past responses and INGA~\cite{loeser2005semantic} routes toward previous query issuers using explicit query topics.
This design leaves owners in control of which documents they share, while peers route queries using document and query embeddings and peer referrals.

The key technical challenge \sys solves is effective routing and overlay management without global knowledge or coordination.
Rather than consulting a centralized directory or broadcasting queries to all peers, a query in \sys involves a best-first walk over the overlay that adapts as queries flow through the network while also progressively rewiring peer connections toward semantically relevant neighbors.
The result is a retrieval layer for \ac{RAG} whose recall improves over time as the overlay organizes, scales gracefully with the number of peers and documents in terms of communication and computation overhead, and remains resilient to peer churn, without requiring network-wide maintenance.

We implement \sys and evaluate its efficiency and effectiveness for decentralized document retrieval in \ac{RAG} pipelines, using three real-world workloads (\aolia~\cite{Pass2006Picture,MacAvaney2022Reproducing}, \amazon~\cite{amazonDataset}, and \pubmedqa~\cite{jin2019pubmedqa}) and two decentralized baselines (\semantica~\cite{semantica} and \dhtlsh).
Our experimental evaluation demonstrates that \sys consistently outperforms other decentralized baselines while requiring no globally maintained structure.
On the \aolia workload, \sys achieves this at one tenth of the message overhead of the strongest baseline (\dhtlsh). 
\sys approaches the recall and end-to-end \ac{RAG} answer accuracy of a centralized document store, without any global index, and 99\% of queries locate the best supporting document in at most 13 peer contacts.
Beyond static performance, our ablation studies show that \sys self-heals after catastrophic failure in which 50\% of all peers crash simultaneously, quickly adapts as new topics are introduced, and scales to networks of up to one hundred thousand peers and documents.
We make the following contributions:
\begin{itemize}
    \item We introduce \sys, a decentralized routing layer for \ac{RAG} that enables document retrieval by routing queries across a decentralized overlay without labeled data, shared data structures, or centralized orchestration (\Cref{sec:zerorag-design}). 
    \item We evaluate
    three real-world datasets, comparing 
    against centralized and decentralized baselines on recall, \ac{RAG} answer accuracy, message efficiency, resilience to peer churn, and topic introduction (\Cref{sec:experiments}).
    We show
    that \sys matches or exceeds the performance of decentralized alternatives while scaling gracefully to large peer populations.
\end{itemize}

\begin{figure*}[t]
    \centering
    \includegraphics[width=.98\linewidth]{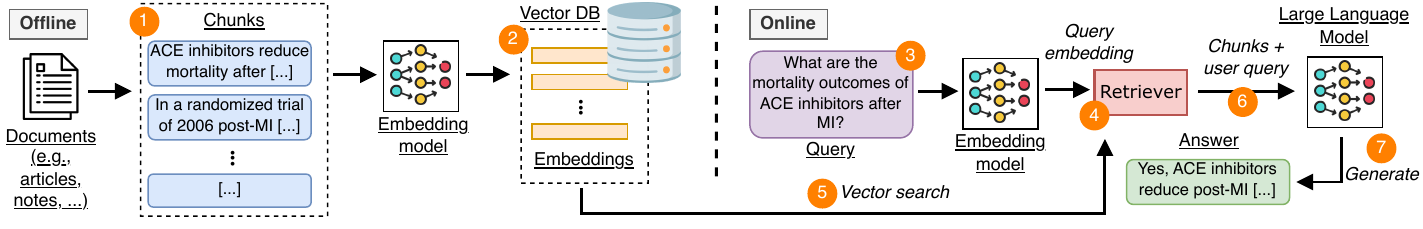}
    \caption{The standard \ac{RAG} pipeline: offline indexing (left) encodes documents into a vector database; online retrieval (right) matches a query to relevant chunks before \ac{LLM} answer generation.}
    \label{fig:rag}
\end{figure*}

\section{Background and Problem Description}
\label{sec:background}
Our aim
is to construct a decentralized document retrieval layer for \ac{RAG} pipelines.
We first describe the standard \ac{RAG} workflow and then discuss the technical challenges that this work addresses.

\subsection{Centralized \ac{RAG} Workflows}
\Acf{RAG} is a standard and popular approach to improve the factual reliability of \ac{LLM} outputs by supplying external context relevant to the user query directly in the prompt~\cite{lewis2020retrieval}.
Whereas conventional language models depend entirely on parametric knowledge encoded during pre-training, \ac{RAG} consults an external data source at inference time, yielding more accurate, up-to-date responses and reducing hallucinations without requiring costly retraining or finetuning.
We illustrate the \ac{RAG} workflow in \Cref{fig:rag}, using an example where a user asks a medical question, and explain the individual operations next.

The pipeline operates in two phases.
In the offline phase, documents, videos, or other source material are preprocessed: each item is divided into chunks, and every chunk is encoded into a high-dimensional vector by an embedding model~\cite{reimers2019sentence}~{\color{orange}\ding{202}}, with the resulting embeddings stored in a vector database~{\color{orange}\ding{203}}.
For presentation clarity, we will refer to chunks as documents throughout this work.
In the online phase, a user query~{\color{orange}\ding{204}} is embedded and forwarded to the retriever~{\color{orange}\ding{205}}, which locates the most semantically similar stored vectors~{\color{orange}\ding{206}} and fetches the corresponding documents from the vector database~{\color{orange}\ding{207}}.
The original query and the retrieved context are then jointly passed to the \ac{LLM}, which produces a response grounded in that evidence~{\color{orange}\ding{208}}, yielding a more accurate and context-aware answer than generation without retrieval. 

The vector database is the central component enabling retrieval. 
Rather than matching exact keywords, it 
uses 
similarity search over high-dimensional embedding spaces.
Prior to insertion, each document is encoded by a pre-trained \ac{ML} model,
typically a sentence transformer~\cite{reimers2019sentence}.
Encodings
allow the database to capture semantic relationships between queries and documents, supporting 
retrieval beyond what keyword-based indices permit~\cite{ma2023comprehensive}.
Incoming queries are
embedded in the same space, and the database returns the nearest neighbors 
using
vector distance.
For tractability on
large corpora, \ac{ANN} indexing techniques are 
used
to accelerate this search~\cite{li2019approximate}.

\subsection{Data Ownership and Availability} 
\label{sec:pitfalls}
Our objective is to build a decentralized mechanism for the retrieval of relevant documents.
A vast body of works exists surrounding peer-to-peer technologies and information retrieval~\cite{risson2007survey}.
Our work challenges two 
commonplace
assumptions 
of these
technologies and \ac{RAG}: (1) that data can be freely ingested regardless of copyright or access control, and (2) that data must be globally available.


\textbf{Copyright and access control.}
Both peer-to-peer retrieval systems and \ac{RAG} architectures typically assume that information is generally available; \ac{RAG} systems in particular oftentimes expect to freely ingest any corpus.
We argue that this assumption will collapse under scrutiny when it comes to the intended access control or copyright law~\cite{rosati2025infringing}.
\Acp{LLM} have been shown to be trained on scraped websites, whereas their terms of service prohibit this, and some even on pirated material, completely circumventing access control~\cite{maini2024llm}.
There have also been various court cases, that were lost by the companies behind the technology, where \ac{ML} algorithms generated copyrighted material verbatim~\cite{amjad2026ai}.
Strict access control is therefore a hard requirement in \ac{RAG} systems, particularly in 
domains where data sovereignty is key, like healthcare or finance.
By adhering to purely decentralized interactions,  \sys is compatible with existing access control mechanisms and peers can refuse or allow other peers to access their information at their discretion.

\textbf{Global data availability.}
Major decentralized systems, \eg, \textsc{Bitcoin}~\cite{nakamoto2008bitcoin}, \textsc{BitTorrent}~\cite{pouwelse2005bittorrent}, and \textsc{IPFS}~\cite{daniel2022ipfs}, are built around the assumption that data should be available everywhere, and they solve for this by replicating data across nodes.
As previously discussed, replication is not always legal or permissible: a hospital, for instance, cannot freely copy patient records to arbitrary peers. 
Data has to stay with the peer that owns it
and peers make no guarantees about availability or access.
Rather than solving the replication problem, \sys 
routes
queries to find data where it already resides, using only local queries and data set descriptions, without 
replicas
of the underlying documents 
in the network.
This comes at a cost: because \sys maintains no global index, it cannot offer the 
routing-time guarantees that a system like \textsc{Kademlia}~\cite{kademlia} provides, and retrieval is therefore best-effort by design.

\subsection{Towards Decentralized \ac{RAG}}

Our routing protocol should, given a user query, locate non-local owners of target or relevant documents to that query under bounded coordination and without centralized control.
Forwarding queries 
to retrieve 
documents is a promising approach to decentralizing \ac{RAG} workflows.
This allows peers to perform 
access control on three levels: identity-based (or attribute-based) access control, query-based access control, and view-based (or purpose-based, or context-aware) access control.
Our innovation therefore is in the \emph{Retriever} component in the \ac{RAG} workflow (step~{\color{orange}\ding{205}} in \Cref{fig:rag}), which still receives results from the local vector database, but also combines these with results from (remote) peers in an overlay network.
This introduces two new challenges: (1) \sys must quickly find and query peers in the network that hold relevant documents to a given query, and (2) \sys must adapt to changes in the available documents over time, \eg, due to peers joining and leaving or updating their local stores, causing semantic drift 
when
the topology becomes misaligned with the current distribution of documents and queries.

\textbf{Challenge 1: locality-aware peer sampling.}
Simply querying all peers in a decentralized overlay 
ensures
maximum recall in terms of document retrieval, but 
cannot work on Internet-scale 
networks
or data volumes.
Instead, each peer should maintain connections to a small set of semantically proximate peers, such that relevant documents can be retrieved from currently connected neighbors alone, without sacrificing recall.
This is the goal of \emph{locality} optimization: steering peer connections toward neighbors whose document collections are likely to satisfy incoming queries~\cite{voulgaris2004exploiting}.
For example, in a network of hospitals with different specializations, two pediatric hospitals are more likely to find similar cases in each other's documents than in those of hospitals with unrelated specializations. However, in semantically homogeneous settings, such as a network of \emph{only} pediatric hospitals, all peers hold similar documents, so any neighbor is as good as any other and a locality-aware overlay brings no benefit over random connections. Therefore, we assume a heterogeneous setting.
However, locality optimization cannot be treated as a one-time configuration: peers may leave the network, and the documents they hold may change over time, shifting what constitutes a relevant neighborhood.
The first challenge addressed by \sys is \emph{maintaining temporally adaptive peer connections that remain semantically relevant} as the network evolves.


\textbf{Challenge 2: connection reconfiguration.}
Existing decentralized retrieval systems that achieve efficient lookup typically do so by maintaining a globally shared index structure, such as a semantic trie~\cite{semantica} or a \ac{DHT}~\cite{kademlia}.
These structures require continuous network-wide reorganization as peers join and leave and as document collections evolve.
In our setting, where coordination must remain strictly local, maintaining such a global structure is not an option.
Thus, the second challenge that \sys takes on is to \emph{reconfigure neighborhood connections directly} as queries and documents evolve.
Our approach is comparable to how \ac{BGP} routes packets without a global directory: each router maintains only local routing tables that are updated incrementally as traffic flows through the network, and efficient paths emerge from these local updates alone.



\section{ZeroRAG Architecture and Design}
\label{sec:zerorag-design}

\textsc{ZeroRAG} is a decentralized retrieval layer that routes queries to peers that hold relevant documents.
In this section, we first discuss the actors and assumptions, and then peers' local algorithms for neighborhood maintenance and query answering.

\begin{figure*}[t]
    \centering
    \includegraphics[width=.92\linewidth]{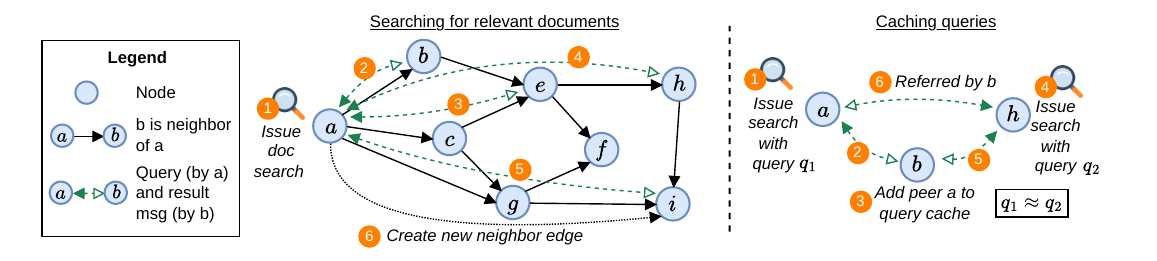}
    \caption{\emph{(left)} Example of 4 query steps of a document searched by peer $a$ with query $q$. After the 
    search concludes, peer $a$ adds peer $i$, which has the most relevant document to $a$'s query, as neighbor. \emph{(right)} The dynamics of query caching in \sys. 
    }
    \label{fig:zerorag}
\end{figure*}

\subsection{System Model}
%
%

Our system model focuses on query routing; we empirically evaluate the accuracy of the downstream \ac{RAG} answer, but do not model the quality of the citation or the reduction of hallucinations. 

\textbf{Network and communication.} \sys coordinates decentralized document retrieval by maintaining a permissioned decentralized overlay network and reconfiguring connections between peers when queries are issued.
The incoming and outgoing connections by each peer form a directed graph.
Peers joining the \sys network can bootstrap into the network by connecting (outgoing) to several random peers, which can be obtained by using existing peer sampling services~\cite{jelasity2007gossip,guerraoui2024peerswap}.
Threats targeted at the network level, like Sybil~\cite{douceur2002sybil} and Eclipse~\cite{heilman2015eclipse} attacks, are beyond the scope of this paper, but we note that peer sampling mechanisms can overcome these after a cold-boot period of several minutes~\cite{STOKKINK2023109701}.

\sys only uses primitive request-response messages (presented shortly), does not maintain shared data structures, or require synchronization steps between peers.
Therefore, messages may be asynchronous or drop.
Likewise, peer connections can drop at any time, \eg, when a peer goes offline or crashes.
We assume the peer sampling mechanism can detect
dropped connections 
and that this
can be implemented with minimal overhead using a ping-pong scheme, message-reception acknowledgments and timeouts.
Beyond this, we do not investigate Byzantine behavior of peers, like attempts to bias others' search results or black hole attacks~\cite{abdelhamid2023lightweight}.


\textbf{Documents and embeddings.}
In \sys, peers exchange embeddings of documents and queries.
The embeddings are used for indexing and network reconfiguration.
The transfer of the documents corresponding to those embeddings is outside \sys.
Exchanged embeddings need to be stored by peers and each peer operates a local vector database for this.
We assume that peers use the same embedding model $\phi_e$, so embeddings produced by different peers remain comparable.
As embeddings are local, 
routing does not depend on centrally assigned labels or a global topic taxonomy, which was the common approach in the late 2000's~\cite{loser2004taxonomy,shen2006building}.

Peers are expected to accrue documents with attribution over time, \ie a verifiable record of which peer a document originates from, as they exchange them after \sys matches their embeddings to queries. 
The attribution itself is out of the scope of \sys and can be implemented with a digital signature.
Within the scope of the paper we are interested in routing, not in the manner in which this history is built.
Therefore, we shortcut the emergent effect of collecting documents by using a \texttt{HISTORY()} remote procedure call, which gathers the history of accrued documents from neighbors \textit{in memory}.  
When considering deployment of \sys, this can (and should) be replaced with 
a performant storage layer.

\subsection{High-Level Description}
\sys is built around the insight that if a peer's neighbors hold semantically similar content, then a query is best served by greedily walking toward the most similar peer at each step.
To this end, whenever a peer $i$ connects to another peer $j$, $j$ sends a set of \emph{advertisements} to $i$.
These advertisements are compact summaries, in the form of document embedding vectors, of the content $j$ locally holds and wants to advertise for routing. This allows owners to retain their documents and decide whether and what to disclose, which is not the same as privacy-preserving retrieval: advertisements, query embeddings, and issuer referrals may reveal information even when documents are withheld.
Now, some peer $p$ aims to obtain the top-$k$ documents that are relevant to a query $q$.
To facilitate this search, peer $p$ maintains a \emph{frontier}, which is a ranked list of candidate peers to visit, ordered by their estimated relevance to the query.
The best-first walk proceeds by repeatedly popping the highest-scoring peer from the frontier, sending it the query, and collecting its response: a set of relevant local document embeddings together with new \emph{referrals} to that peer's own neighbors.
These referrals are scored against the query and put back into the frontier, extending the search into previously unvisited parts of the network.
Meanwhile, peer $p$ keeps track of the document embeddings that it has seen.
After a fixed number of steps, $p$ will inspect the 
these
embeddings 
and compare their relevance against query $q$.
Peer $p$ 
then retrieves the top-$k$ relevant documents 
from the peers that offer them.

\sys continuously reconnects the overlay itself as queries are issued.
After each query completes, the querier adds a directed edge toward whichever peer returned the most relevant result, evicting the oldest existing connection if the neighbor list is full.
This implicit feedback loop causes the overlay to self-organize around semantic affinity: peers that frequently serve useful results for a given topic accumulate more incoming connections, and future queries on that topic are naturally steered toward them earlier in the walk.
As the overlay network reconnects, peers are increasingly likely to quickly find the documents they need, without any global coordination or shared data structure.
In addition, each peer keeps a bounded routed-query memory: a cache of embedding vectors from queries it has recently responded to, which were issued by other peers.
When scoring an incoming query, a peer consults this memory alongside its neighbor advertisements, and can 
refer the querier not only toward peers with relevant documents but also toward peers that have recently asked similar questions, and are thus likely to be connected to, or themselves hold, relevant content.

\textbf{Illustrative example.}
\Cref{fig:zerorag} (left) shows the progression of a document search, issued by peer $a$~{\color{orange}\ding{202}}.
Peer $a$ iteratively sends search queries to peers in its frontier and receives 
replies with relevant documents and referrals.
We show a document search where four peers are queried by $a$~({\color{orange}\ding{203}}-~{\color{orange}\ding{206}}).
After the search completes, peer $a$ adds peer $i$ as neighbor since it has the most relevant documents~{\color{orange}\ding{207}}.

\Cref{fig:zerorag} (right) shows query caching.
Peer $a$ issues a search with query $q_1$ and queries peer $b$~({\color{orange}\ding{202}},{\color{orange}\ding{203}}).
Peer $b$ adds peer $a$ and the incoming search query to its local query cache~{\color{orange}\ding{204}}.
When another peer $h$ later issues a query $q_2$, which is similar to $q_1$ (\ie, their embeddings are close) and interacts with peer $b$, $b$ refers peer $h$ to peer $a$ as $a$ is likely to have found relevant documents for $q_2$.

\begin{algorithm}[t]
\caption{Retrieving top-$k$ documents for query $q$ 
of peer $u$.
}
\label{alg:zerorag_querier}\small
\begin{algorithmic}[1]
\Require $C$: contact budget, $r$: referral limit, $\delta_n$: max amount of neighbors each peer maintains
\Require Persistent state $\sigma_u = (D_u, N_u, A_u, M_u)$ \Comment{
    local documents $D_u$, outgoing neighbors $N_u$, neighbor advertisements $A_u$,
    and routed-query memory $M_u$.}
\Procedure{retrieveDocuments}{$q$: query, $k$: result limit}
    \State $e_q \gets \phi_e(q)$ \Comment{generate embedding of $q$}
    \State $Q \gets \emptyset$ \Comment{peers already contacted}
    \State $R \gets \emptyset$ \Comment{document results found so far}
    \State $F \gets \{(u, -\infty)\}$ \Comment{frontier of peers to contact}
    \State $B \gets \emptyset$ \Comment{best referral score seen for each peer}
    \While{$F \neq \emptyset$ \textbf{and} $|Q| < C$}
        \State $(x, \_) \gets \arg\max_{(v,\, s) \in F}\, s$ \Comment{peer with highest score}
        \State $F$.\textsc{removePeer}($x$)
        \If{$x \in Q$}
            \State \textbf{continue}
        \EndIf
        \State $Q \gets Q \cup \{x\}$
        \State \textbf{send} \textsc{query}$(e_q, k)$ message to peer $x$
        \State \textbf{receive} \textsc{queryResult}$(H_x, \mathit{Refs}_x)$ message from peer $x$
        \State $R \gets \textsc{topK}(R \cup H_x, k)$
            \Comment{keep the best document results}
        \ForAll{$(v, s) \in \mathit{Refs}_x$}
            \If{$v \notin Q$ \textbf{and} $(B[v]$ is undefined \textbf{or} $s > B[v])$}
                \State $B[v] \gets s$
                \State $F$.\textsc{update}($(v,s)$)
            \EndIf
        \EndFor
    \EndWhile
    \State \textsc{updateNeighbors}($R$, $B$) \Comment{Overlay maintenance}
    \State $D \gets$ \textsc{obtainDocuments}($R$) \Comment{retrieve docs from docrefs}
    \State \Return $D$
\EndProcedure
\vspace{.2\baselineskip}
\Procedure{updateNeighbors}{$R$: found docrefs, $B$: ref scores}
    \State $a \gets \bot$
    \If{$R \neq \emptyset$} \Comment{We found at least one document}
        \State $a \gets$ peer returning the highest-scoring document in $R$
    \ElsIf{$B \neq \emptyset$}
        \State $a \gets \arg\max_v B[v]$
    \EndIf
    \If{$a \neq \bot$ \textbf{and} $a \neq u$}
        \State $N_u$.\textsc{add}($a$)
        \If{$|N_u| > \delta_n$}
            \State $N_u$.\textsc{evict}() \Comment{Evict oldest neighbor}
        \EndIf
    \EndIf
\EndProcedure
\end{algorithmic}
\end{algorithm}

\subsection{Technical Description}


We describe the concrete steps in \sys through the roles of a peer $u$ that aims to retrieve $k$ documents relevant to a query $q$ (with embedding $e_q$) by repeatedly sending \textsc{query} messages to candidate peers in the frontier, and a peer $p$ that receives such a \textsc{query} message.
These roles are purely explanatory: any peer can fulfill either role, even at the same time.
\Cref{alg:zerorag_querier} explains the search procedure from the perspective of peer $u$ and \Cref{alg:zerorag_handleQuery} shows how peer $p$ handles an incoming query.

\textbf{Local state.}
To facilitate routing
, every peer $i$ stores a local state $\sigma_i$, given by the 4-tuple $(D_i, N_i, A_i, M_i)$.
The documents stored in the local database of peer $i$ are represented by $D_i$.
The neighbors of $i$ are given by $N_i$, which is a FIFO list with limited capacity $\delta_n$.
Insertion and eviction from this list is based on the most recent highest-scoring query.
The collection of advertisements collected by $i$, of its neighbors, is given by $A_i$.
Lastly, the collection of the most recent $\delta_m$ received pairs ($u$, $e_q$), as a result of acting as the role of the queried peer $p$, form the chronological FIFO list $M_i$. The local state is designed for the simplest effective implementation. FIFO avoids maintaining estimates of each neighbor's usefulness: eviction needs only insertion order~\cite{voulgaris2004exploiting}.
Maximum scoring allows one close match to guide routing, despite unrelated vectors~\cite{tang2003peer}.
One neighbor per query limits connection changes and advertisement exchanges~\cite{sripanidkulchai2003efficient}.

Both algorithms \ref{alg:zerorag_querier} and \ref{alg:zerorag_handleQuery} use \textit{scores} of relevance.
These scores are simply the cosine similarity of two embeddings, or a score of minus infinity if the two have not been compared.
\sys attempts to localize documents with the highest score in the network.

\textbf{Retrieving documents.}
A peer $u$ invokes the \texttt{retrieve\-Doc\-uments} procedure, given in \Cref{alg:zerorag_querier}, to retrieve up to $k$ of the most-relevant documents to the query $q$.
The remainder of the algorithm will use the embedding $e_q$ of $q$ (line 2) to query neighbors of $u$.
If the required documents are not found locally, the peer constructs a frontier $F$ of uncontacted neighbors, and successive neighbors of neighbors.
It initially consists of pairs of currently-connected neighbors $N_u$ of $u$ and scores of minus infinity to signify that they have not been queried (line 5).
During the following best-first-search loop (line 7), $F$ may already receive scores for identities from previously queried peers.
The search loop successively visits any (there may be multiple with the same score) most high-scoring peer $x$ in the frontier, up to $C$ peers, inserting $x$ into the visited peer set $Q$ (line 7-13).
Peer $u$ then sends a \textsc{query} message to peer $x$, containing the query embedding $e_q$ and the value of $k$ (line 14).
Peer $u$ awaits a response \textsc{queryResult} message from $x$, containing the top remote results and top referrals of $x$ (line 15).
The document results are merged into the set $R$, which is then trimmed down to the top-$k$ (line 16).
Returned identities and their scores are inserted into $F$, or the identity's score is updated if it was already present, while the best-known score and its (potentially unvisited) associated identity is saved in  $B$ (lines 17-22).
After the search loop is finished, the network connections are reconfigured (line 24) and the documents are retrieved, by their embedding, from the peer that stores them and returned (line 25 and 26).


\textbf{Neighborhood reconfiguration.}
As a result of a query and its retrieved document embeddings $R$ and best identity-score pair $B$, the current neighbors of $u$ are updated using the \texttt{updateNeighbors} procedure, given in \Cref{alg:zerorag_querier} (line 28).
The identity $a$ of the peer that has the highest-scoring document embedding in $R$ is retrieved (line 31).
Otherwise, if $R$ was empty, the identity of the peer $a$ that was advertised as the best score is retrieved (line 33), if there was one.
If an identity $a$ is available, it is connected to and if that causes the current number of neighbors to exceed $\delta_n$ the oldest neighbor is removed from the list (lines 35-40).


\textbf{Query handling.}
We next explain how peer $p$ handles an incoming \textsc{query} message, also see \Cref{alg:zerorag_handleQuery}.
First, peer $p$ queries its local vector database of to retrieve document embeddings $H$ based on $e_q$ (line 2).
Then, $p$ forms a pool of routable, known embeddings $E$ by all known pairs of document embeddings and query embeddings of neighbors, coupled to the identity of the neighbor identifier $v$ that sent them (line 3).
The scores $S$ of identities are initialized with minus infinity (line 4) and then set to the highest score over the elements in $E$ that are associated with it (lines 5-9).
The top $r$ peers are then taken from $S$ (line 10) to be returned as the referrals $Refs$ later.
The pair of the querying peer $u$ and its query embedding $e_q$ is stored in the routed-query memory $M_p$ (line 11), to support future routing of queries similar to $u$.
The oldest entry in $M_p$ is evicted if the number of these tuples exceeds $\delta_m$ size (lines 12-14).
Finally, $p$ sends a \textsc{queryResult}, with the local top-$k$ embeddings in $H$ and the top-$r$ referrals in $Refs$ (line 15).


\textbf{Retrieval parallelization.}
To reduce end-to-end 
retrieval latency, it is possible to parallelize 
\Cref{alg:zerorag_querier}.
Instead of performing a best-first search, a breadth-first search can 
send and await
up to  $f$ \textsc{query} and \textsc{queryResult} messages in parallel.
As requests and responses can be asynchronous, the number of visited peers that have responded might then overshoot the contact budget $C$ by the node fanout of the connection graph.
In the remainder of this paper, we use the sequential version of \Cref{alg:zerorag_querier} to avoid additional variance in our metrics, which do not include wall-clock time.

\textbf{Cleanup.} Our algorithms do not show the response of \sys to churn, as signaled by the peer sampling.
When peer $p$ leaves, peer $u$ removes $p$ from $N_u$ and deletes entries associated with $p$ from $A_u$ and $M_u$.
Documents in $D_u$ 
do
not need 
cleanup and other peers in the network will now start being routed to $u$ instead of $p$.

\sys only has a \textsc{query} and a \textsc{queryResult} message that can be dropped.
The dropping of either of these can be remedied by waiting until some timeout when receiving the response message in \Cref{alg:zerorag_querier} (line 15).
If the timeout is triggered, the query results are an empty list of documents and an empty list of referrals.

\textbf{Scalability and topology convergence.}
With fixed-size encodings, each query sends at most \(C\) requests and \(C\) replies, carrying \(O(C((k+1)d+r))\) bytes for \(d\)-dimensional embeddings, \(k\) results and \(r\) referrals. If each neighbor advertises at most \(a\) vectors, removed when that neighbor is evicted, persistent routing state occupies \(O(\delta_n+d(\delta_n a+\delta_m))\) bytes, where \(\delta_n\) and \(\delta_m\) cap neighbors and cached queries. Each search adds at most one connection, transferring \(O(ad)\) advertisement bytes. 
The total routing state and traffic for one query per peer are \(O(n)\) for \(n\) peers, excluding document transfer, retries, and background maintenance. 
Regarding topology convergence, stability can only be achieved if the number of connected peers matches the number of interests. At capacity \(\delta_n\), adding a new neighbor for the current query evicts the oldest connection, potentially displacing a peer useful to other interests; in this configuration therefore the topology cannot feasibly converge.

\begin{algorithm}[t]
\caption{Query handling at peer $p$.}
\label{alg:zerorag_handleQuery}\small
\algtext*{EndUpon} 
\begin{algorithmic}[1]
\Require $r$: referral limit, $\delta_m$: memory size limit
\Require Persistent state $\sigma_p = (D_p, N_p, A_p, M_p)$

\Upon{\textsc{Query}$(e_q, k)$ message \textbf{from} peer $u$}
    \State $H \gets D_p$.\textsc{search}$(e_q, k)$ \Comment{local document search}
    \State $E \gets A_p \cup M_p$
        \Comment{advertisements and past routed queries}
    \State $S[v] \gets -\infty$ \textbf{for} every $(v, \_) \in E$
    \ForAll{$(v, z) \in E$} \Comment{compute best scores for peers}
        \If{$v \neq u$}
            \State $S[v] \gets \max\{S[v],\, \langle e_q, z \rangle\}$
        \EndIf
    \EndFor
    \State $\mathit{Refs} \gets \textsc{topK}(\{(v, S[v]) : S[v] > -\infty\},\, r)$
    \State $M_p$.\textsc{add}($(u, e_q)$)
    \If{$|M_p| > \delta_m$}
        \State $M_p$.\textsc{evict}() \Comment{Evict oldest routed-query memory item}
    \EndIf
    \State \textbf{send} \textsc{QueryResult}$(H, \mathit{Refs})$ message to peer $u$
\EndUpon
\end{algorithmic}
\end{algorithm}

\section{Experimental Evaluation}
\label{sec:experiments}


We implement \sys as routing middleware between a network overlay and local document stores.
Our experimental evaluation answers the following five questions:

\begin{enumerate}
\setlength{\itemsep}{0pt}\setlength{\parskip}{0pt}
    \item What is the performance of \sys on real-world data sets in terms of recall and accuracy (Section \ref{sec:exp-aol})?
    \item How does the end-to-end accuracy of RAG change when using \sys (Section \ref{sec:expe2e})?
    \item How resilient is \sys to network-wide failure (Section \ref{sec:expchurn})?
    \item What is the efficacy of \sys's inputs and what is the effect of their ablation (Section \ref{sec:ablation})?
    \item How quickly does \sys make queries routeable when documents with new topics are introduced (Section \ref{sec:expdrift})?
\end{enumerate}


%
%

\subsection{Experimental Setup}
\label{sec:method-datasets}

We next describe our experimental setup, including our workloads, baselines and implementation details.

\subsubsection{Workloads}
We evaluate \sys and baselines using three real-world workloads: \aolia~\cite{Pass2006Picture,MacAvaney2022Reproducing}, \amazon~\cite{amazonDataset}, and \pubmedqa~\cite{jin2019pubmedqa}. 
The three workloads cover complementary retrieval settings: single-target and multi-relevant retrieval, real and synthetic user behavior, and general-domain, e-commerce, and biomedical query distributions.
We additionally use \pubmedqa for an end-to-end \ac{RAG} evaluation with \sys, extending our evaluation beyond routing performance alone.

\textbf{\aolia} provides a query-and-click stream with user ids, query text, clicked documents, and timestamps.
The full workload contains \num{657426} users, \num{428157} documents and \num{2594705} total queries.
We adapt the workload to our setting by selecting the \num{10000} users with the most queries, together performing \num{2171460} queries.
Prior to running this workload, we populate the local document stores with the documents that the user has clicked.
During our evaluation, we then replay these events in a randomized order by having peers search for the document they are interested in (\eg, $k=1$), without them looking into their local database first.
The single-target retrieval objective (find the exact document) enables us to test the effectiveness of query routing in \sys with a binary objective score (whether the document is found or not).

\textbf{\amazon} is an adaptation of the personalized product-search benchmark introduced by Ai et al.~\cite{amazonDataset}.
This benchmark contains \num{63001} product descriptions, and each product contains, besides some basic metadata, a category path (\eg, ``Camera, Photo $\rightarrow$ Digital Camera Lenses'').
The workload authors synthetically constructed search queries by taking each product's category path and flattening it into a query string, stripping stopwords and duplicates.
For example, ``Camera, Photo $\rightarrow$ Digital Camera Lenses'' becomes the query ``photo digital camera lenses.''
Thus, queries are categorical and fairly repetitive across users.
The queries were then mapped to users using the reviews of purchased products: if a reviewer purchased a product associated with a category-derived query, 
a corresponding
reviewer--query pair is made.


In our evaluation, each reviewer in the \amazon benchmark becomes a peer in the network.
Each reviewer–query pair becomes one simulation event, \ie, one search query issued by one peer.
The relevant products for that reviewer–query pair become the documents that peer is searching for.
The retrieval objective for \sys is to recover as many relevant products as possible from other peers, rather than to recover a single clicked item.
Those documents are distributed across other peers in the network, based on which peers reviewed those products.
Because queries come from product categories rather than individual user intent, the same query string will naturally recur across many different reviewers.
This means the experiment ends up simulating a setting of repeated, overlapping product-search demand, where many peers are asking semantically similar things at different times.



\textbf{\pubmedqa} is a biomedical question-answering dataset built from abstracts in the \pubmed database~\cite{jin2019pubmedqa}.
We adopt the discrete-answer subset of \pubmedqa with \num{1000} items, in which each item contains a biomedical question, an abstract, and a categorical answer (\texttt{yes}, \texttt{no}, or \texttt{maybe}).
To increase query-level semantic variety, we supplement each original question with five rephrased queries generated using the \textsc{GPT-5-nano} model, resulting in six query variants (the original question and five rephrasings).


To assign \pubmed documents to peers, we use \ac{MeSH} terms, assigned to \pubmed articles, as biomedical topic labels.
We sample a \ac{MeSH}-level preference vector $\boldsymbol{\theta}_u$ from a $\operatorname{Dirichlet}(\boldsymbol{\alpha})$ distribution, 
where \(\boldsymbol{\theta}_u\) defines user \(u\)'s preference distribution over all \num{23904} \ac{MeSH} topics.
This produces 
user profiles that are biased toward a small number of topics.
We fix $\boldsymbol{\alpha} = 7 \times 10^{-5} $, 
where a peer gets on average 15.9 documents on the first topic, 6.6 on the second topic, 3.4 on the third topic and so on.
Given a target number of documents $t$, we 
sample the number of documents assigned to each \ac{MeSH} topic according to \(\boldsymbol{\theta}_u\).
We fix $t=30$ unless otherwise stated.
We then uniformly sample documents from each selected \ac{MeSH} topic, and for each 
document uniformly choose one of its six query variants. 
Under this user-model, most peers are interested in 1-3 main topics with the interest 
dropping off for other topics.
Each document 
of
a peer $i$ also carries one randomly selected query variant.
This (peer, document, query) triple defines a simulation event: peer $i$ issues the query, and the retrieval objective is to locate the corresponding document on another peer.
Each peer 
issues exactly as many queries as it holds documents.
Unless otherwise stated, experiments with the \pubmedqa workload contain \num{50000} simulated peers and \num{30} queries per peer, leading to $1.5$ million query-document events, issued in random sequence.

\subsubsection{Implementation}

The experiments were simulated by following a timeline of `events', one event representing a peer sending a query which aims to retrieve one or more target documents.
We fix the contact budget $C=50$, meaning that a peer contacts at most 50 peers during document retrieval. 
We used the \textsc{all-mpnet-base-v2} embedding model for embedding generation of documents and queries~\cite{allmpnetbasev2}.
For embedding generation of documents in the \aolia workload, we concatenated the \texttt{title}, \texttt{body}, \texttt{text}, and \texttt{url} values from the data set using the string ``\textbackslash{n}\textbackslash{n}''.
For documents in the \amazon workload, we concatenated their \texttt{title}, \texttt{description}, \texttt{categories}, \texttt{brand}, and \texttt{price} in a similar fashion.

The overlay topology is implemented as a global directed connection graph, 
where each graph node represents a peer.
Message passing between nodes is implemented using remote procedure calls.
Document embeddings are kept in a single unified memory and the local storage of peers are references in that memory.

Figures~3, 4, 6, and~7(b) and 
all
ablation tables use five runs per configuration: 
a
randomized starting topology for each run
of
\sys, and order of the peers in the tree for Semantica.
For matching windows, bins, or table summaries, we compute the five-run mean $\bar{x}$ and sample SD in Figures~4 and~7(b), and the tables.
Figures~3 and~6 instead show the mean with 95\% Student-$t$ confidence intervals; DHT-LSH was not repeated because 
its hash functions 
are
fixed, retrieval does not depend on randomized peer connections, and unrepeated baselines have no uncertainty bars.

\subsubsection{Hardware}
All simulations were performed using 12 computing nodes which are equipped with two AMD EPYC~7282 processors (16 cores per socket, 2.8\,GHz), for 32 physical cores in total.

\subsubsection{Baselines}

We evaluate \sys against two related baselines that together occupy different parts in the design space.

\textbf{\semantica}~\cite{semantica} is the closest prior decentralized semantic retrieval system and organizes peers with a globally maintained semantic trie.
We 
use
the source code of \semantica as published in the author's Github repository. 
For the \ac{HNSW} implementation we utilize \textsc{hnswlib}~\cite{malkov2018efficient}.

\textbf{\dhtlsh} is a baseline that pairs a \ac{DHT} (\eg, \textsc{Kademlia}~\cite{kademlia}) with \ac{LSH} schemes to provide \ac{ANN} search in fully decentralized settings.
\dhtlsh relies on $p$-stable hash functions that map embedding vectors into discrete buckets such that semantically similar vectors collide into the same bucket with high probability~\cite{LSH-pstable-Dist}.
To improve recall, \dhtlsh maintains $L$ independent hash tables, each using a composition of $k$ such functions with bin width $w$.
At indexing time, each document embedding is hashed across all $L$ tables, and the resulting bucket identifiers are used as keys in the \ac{DHT}, which distributes storage of bucket contents across peers.
At query time, the query embedding is hashed through the same $L$ tables, yielding $L$ bucket keys, each of which requires a separate \ac{DHT} lookup to retrieve candidate documents.
We adopt the suggested hyperparameter values from the original LSH paper~\cite{LSH-pstable-Dist}, fixing $k=10$. We adopted $w=3$ instead the of the $w=4$ used by the authors because $w=4$ placed 44.7\% of the AOL-IA documents in the single largest bucket. 

\subsubsection{Metrics}
Our primary success metric is \emph{recall}: the percentage of relevant documents retrieved from the entire set available in the network, excluding documents already held locally by the querying peer.
For \aolia, where each query has a single target document, this reduces to a binary success rate.
For \amazon, where a query may have hundreds of relevant documents, we instead report the \emph{fraction of available relevant documents retrieved} within the routing budget.
For \pubmedqa we additionally measure end-to-end \ac{RAG} accuracy: the fraction of questions answered correctly when the retrieved top-$k$ documents are supplied as context.
This assesses whether routing quality translates into downstream generation quality.
For brevity, in the remainder of this section, we will use the notation \emph{recall@k} for the recall when the retrieved top-$k$ documents are supplied as context.
Unless stated otherwise, we report recall averaged over batches of 1000 queries to smooth per-query variance and surface trends in routing performance as the overlay evolves over time.


\subsection{Effectiveness of Document Retrieval} 
\label{sec:exp-aol}
We first evaluate the effectiveness of document retrieval with different configurations of \sys, and the \semantica and \dhtlsh baselines, using the \aolia and \amazon workloads.
For \dhtlsh, we vary the number of buckets $L \in \{10, 20, 50\}$.

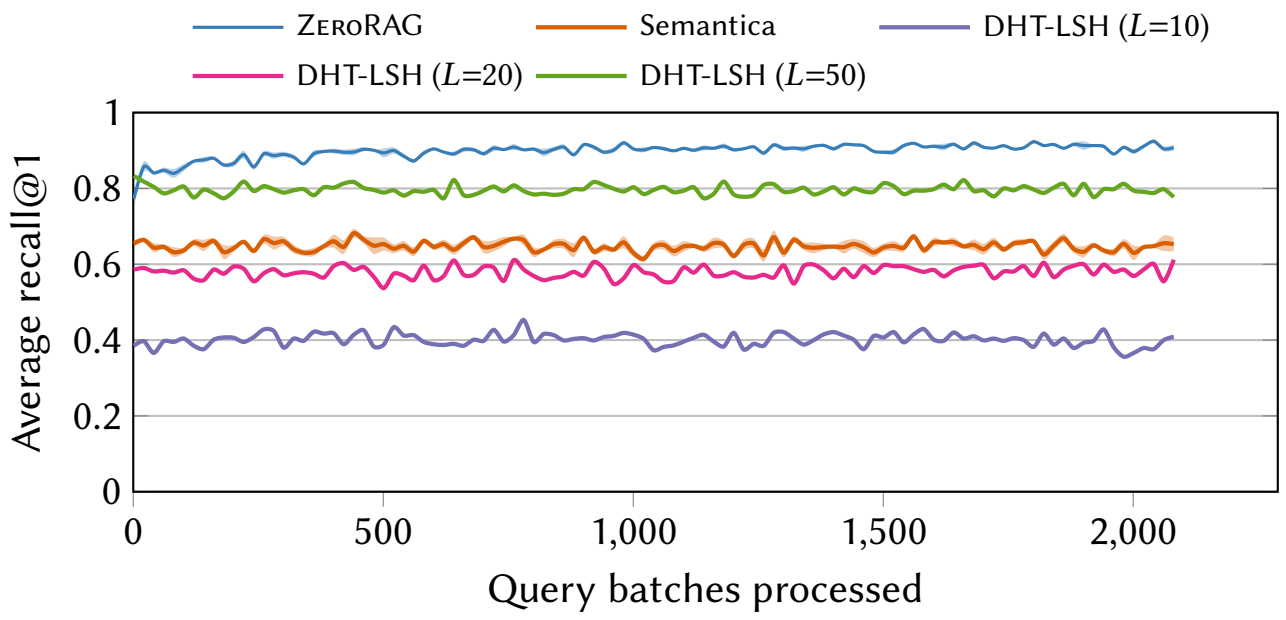
\begin{figure}[t]
    \centering
    \begin{tikzpicture}
  \begin{axis}[
    xlabel={Query batches processed},
    ylabel={Average recall@1},
    ymin=0, ymax=1,
    xmin=0,
    width=260.4pt,
    legend columns=3,
    legend style={font=\sffamily\scriptsize},
    cycle list name=Dark2,
    every axis plot/.append style={thick, no marks, solid, smooth, each nth point=20},
  ]
    \addplot[name path=zrlo, draw=none, forget plot] table[col sep=comma, x=window_index, y={ZeroRAG_ci95_lower}]
      {csv-files/aol_recall_rolling1000.csv};
    \addplot[name path=zrhi, draw=none, forget plot] table[col sep=comma, x=window_index, y={ZeroRAG_ci95_upper}]
      {csv-files/aol_recall_rolling1000.csv};
    \addplot[Set1-B, fill opacity=0.35, draw=none, forget plot] fill between[of=zrhi and zrlo];
    \addplot[name path=selo, draw=none, forget plot] table[col sep=comma, x=window_index, y={Semantica_ci95_lower}]
      {csv-files/aol_recall_rolling1000.csv};
    \addplot[name path=sehi, draw=none, forget plot] table[col sep=comma, x=window_index, y={Semantica_ci95_upper}]
      {csv-files/aol_recall_rolling1000.csv};
    \addplot[Dark2-B, fill opacity=0.35, draw=none, forget plot] fill between[of=sehi and selo];
    \addplot+[sys, semithick] table[col sep=comma, x=window_index, y={ZeroRAG}]
      {csv-files/aol_recall_rolling1000.csv};
    \addlegendentry{\sys}
    \addplot+ table[col sep=comma, x=window_index, y={Semantica}]
      {csv-files/aol_recall_rolling1000.csv};
    \addlegendentry{Semantica}
    \addplot+ table[col sep=comma, x=window_index, y={LSH w=3.0 k=10 L=10}]
      {csv-files/aol_recall_rolling1000.csv};
    \addlegendentry{\dhtlsh ($L{=}10$)}
    \addplot+ table[col sep=comma, x=window_index, y={LSH w=3.0 k=10 L=20}]
      {csv-files/aol_recall_rolling1000.csv};
    \addlegendentry{\dhtlsh ($L{=}20$)}
    \addplot+ table[col sep=comma, x=window_index, y={LSH w=3.0 k=10 L=50}]
      {csv-files/aol_recall_rolling1000.csv};
    \addlegendentry{\dhtlsh ($L{=}50$)}
  \end{axis}
\end{tikzpicture}
    \caption{Average recall@1 for \aolia query batches, 
    for \sys and different baseline configurations.}
    \label{fig:aolia}
\end{figure}

\subsubsection{\aolia workload}
\label{sec:exp_aiolia_workload}
We issue the search queries in the \aolia workload, in randomized order, and measure recall@1, which indicates whether the document has been retrieved or not.
\Cref{fig:aolia} shows average recall@1 for each query batch, for all evaluated systems.
\sys obtains an average recall@1 of 0.91 for the last query batch.
In contrast, this same number for \semantica is 0.65 and significantly lower. 
We attribute this to the fact that peers in \semantica have a single (averaged) embedding that is used to advertise their documents, whereas in \sys peers share more than one non-averaged embedding.
\dhtlsh with $L=50$ approaches \sys with an average recall@1 of 0.80, while \dhtlsh with $L=20$ and $L=10$ reach only 0.56 and 0.41, respectively.
However, the drawback of \ac{LSH} methods is high communication and computation cost.
Every bucket check requires traversing the \ac{DHT}:
the overhead in our \num{10000}-peer network, for instance, is that each search query requires approximately $50 \cdot log_2(10000) \approx 664$ return messages (to another peer and back). 
In contrast, \sys only requires $50$ return messages per search query for similar performance. Furthermore, 
the largest \ac{DHT} bucket for $L = 50$ holds 25\% of all documents, therefore every query routed to that bucket has to be compared against all of them.

\colorlet{shadecolor}{takeawaycolor}
\begin{shaded}
\noindent\textbf{Takeaway (\Cref{fig:aolia}).} Using the \aolia workload, \sys shows superior retrieval effectiveness compared to \dhtlsh and \semantica, at a lower message complexity.
\end{shaded}

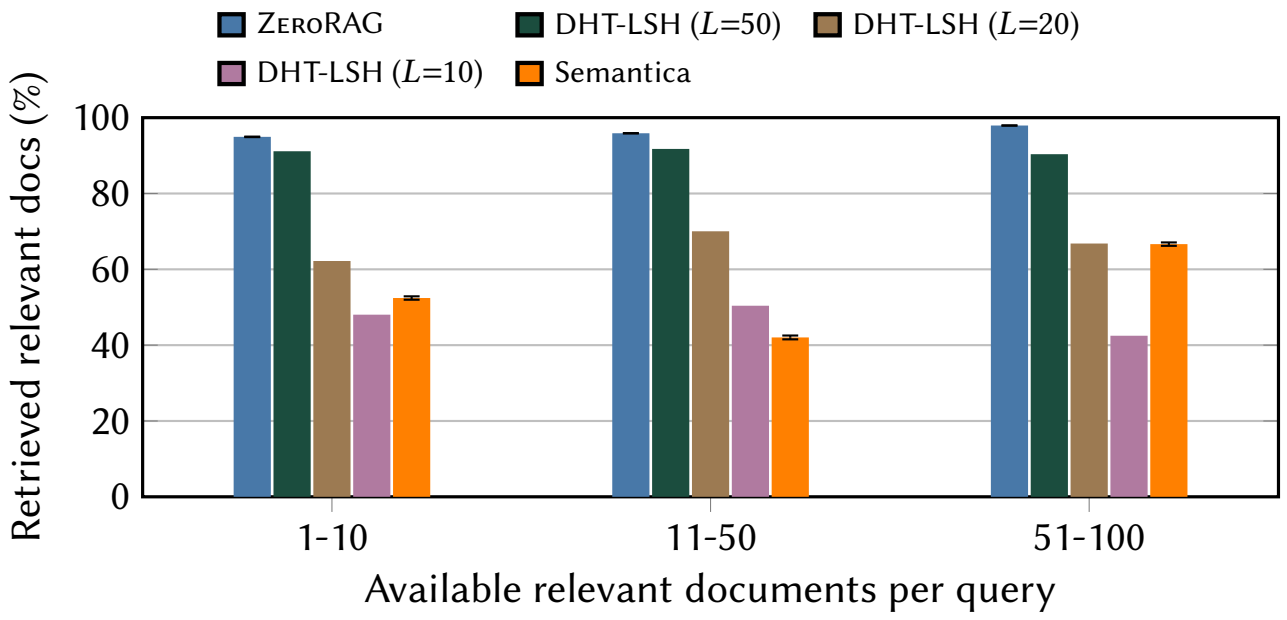
\begin{figure}[t]
    \centering
%
\definecolor{zsblue}{HTML}{4878A8}     
\definecolor{lshbrunswick}{HTML}{1B4D3E}   
\definecolor{lshbrown}{HTML}{9C7A52}   
\definecolor{lshmauve}{HTML}{B07AA0}   
\definecolor{semorange}{HTML}{FF8000}  

\begin{groupwidth}{1}
\noindent\begin{tikzpicture}
  \begin{axis}[
    xlabel={Available relevant documents per query},
    ylabel={Retrieved relevant docs (\%)},
    ymin=0, ymax=100,
    width=258.77pt,
    ybar=0.5pt,
    bar width=7pt,
    symbolic x coords={1-10, 11-50, 51-100},
    xtick=data,
    enlarge x limits=0.25,
    ymajorgrids=true,
    bar legend,
    legend columns=3,
    legend style={
      font=\sffamily\scriptsize,
      at={(0.45,1.02)}, anchor=south,
      /tikz/every even column/.append style={column sep=4pt},
    },
    every axis plot/.append style={fill, draw=none},
  ]
    \addplot[std bars, draw=none, fill=zsblue]     table[col sep=comma, x=relevant_docs_bin, y expr=\thisrow{zerorag_mean_query_retrieval_recall}*100,
        y error expr=\thisrow{zerorag_mean_query_retrieval_recall_std}*100]
      {csv-files/amazon_relevant_retrieval_histogram.csv};
    \addlegendentry{\sys}
    \addplot[draw=none, fill=lshbrunswick]   table[col sep=comma, x=relevant_docs_bin, y expr=\thisrow{lsh_w3p0_k10_l50_mean_query_retrieval_recall}*100]
      {csv-files/amazon_relevant_retrieval_histogram.csv};
    \addlegendentry{\dhtlsh ($L{=}50$)}
    \addplot[draw=none, fill=lshbrown]   table[col sep=comma, x=relevant_docs_bin, y expr=\thisrow{lsh_w3p0_k10_l20_mean_query_retrieval_recall}*100]
      {csv-files/amazon_relevant_retrieval_histogram.csv};
    \addlegendentry{\dhtlsh ($L{=}20$)}
    \addplot[draw=none, fill=lshmauve]   table[col sep=comma, x=relevant_docs_bin, y expr=\thisrow{lsh_w3p0_k10_l10_mean_query_retrieval_recall}*100]
      {csv-files/amazon_relevant_retrieval_histogram.csv};
    \addlegendentry{\dhtlsh ($L{=}10$)}
    \addplot[std bars, draw=none, fill=semorange]  table[col sep=comma, x=relevant_docs_bin, y expr=\thisrow{semantica_mean_query_retrieval_recall}*100,
        y error expr=\thisrow{semantica_mean_query_retrieval_recall_std}*100]
      {csv-files/amazon_relevant_retrieval_histogram.csv};
    \addlegendentry{Semantica}
  \end{axis}
\end{tikzpicture}
\end{groupwidth}
    \caption{Percentage of retrieved results, binned by the number available results for a query, for the Amazon workload.}
    \label{fig:amazon_results_histogram}
\end{figure}

\subsubsection{\amazon workload}

We use the same setup as in the experiment in \Cref{sec:exp_aiolia_workload} but run the \amazon workload.
However, queries in this workload may have many relevant documents rather than just a single result.
Therefore, recall@1 is not well defined and we address this by considering different values of $k$ when computing recall@$k$ for three different bin sizes, based on how many relevant documents there are for each query: if there are 1-10 available relevant documents, we use recall@10, recall@50 for queries with 11-50 available documents, and recall@100 for queries with 51-100 available documents.

\Cref{fig:amazon_results_histogram} shows the fraction of retrieved relevant documents for \sys and baselines, grouped by the number of available documents for each query.
\sys retrieves 94.8-98.9\% of available relevant products across all three bins.
\dhtlsh benefits from broader hash-table coverage but falls behind \sys: with $w{=}3.0$, $k{=}10$, $L{=}50$ it reaches 90.3-91.7\%, with $L{=}20$ being 62.1-70.0\%, while a value of $L{=}10$ drops it to 42.5-50.3\%.
\semantica does match the performance of \dhtlsh $L{=}20$ in the largest bin (66.7\%) but falls behind again in the smaller and middle bins ( 41.7-52.9\%).



\colorlet{shadecolor}{takeawaycolor}
\begin{shaded}
\noindent\textbf{Takeaway (\Cref{fig:amazon_results_histogram}).} 
In terms of recall, \sys consistently outperforms \dhtlsh and \semantica on the \amazon workload,
which has varying numbers of results per query instead of one best result.
\end{shaded}




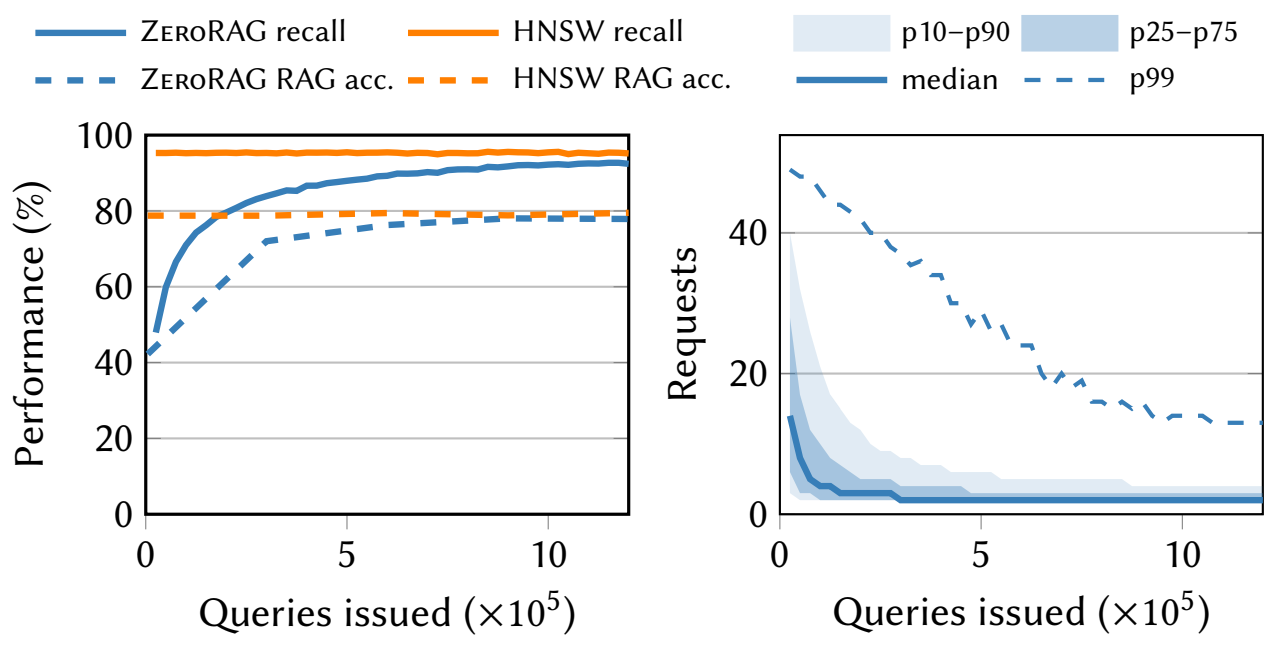
\begin{figure}[t]
    \centering


\begin{tikzpicture}
  \begin{groupplot}[
    group style={
      group size=2 by 1,
      horizontal sep=1cm,
      group name=g,
    },
    width=135.89pt,
    legend columns=1,
    xmin=0, xmax=12,
    scaled x ticks=false,
    xticklabel style={/pgf/number format/.cd, fixed, precision=0},
    every axis plot/.append style={very thick, no marks},
    legend style={font=\sffamily\scriptsize, fill=white, fill opacity=0, text opacity=1,at={(0.5,1.05)}},
  ]
    \nextgroupplot[
      xlabel={Queries issued ($\times 10^{5}$)},
      ylabel={Performance (\%)},
      ymin=0, ymax=100,
      legend columns=2,
      unbounded coords=discard,
    ]
    \addplot+[sys, discard if not={series}{ZeroRAG retrieval top-5 recall}]
      table[col sep=comma, x expr=\thisrow{event_index}/100000, y=value_pct]
      {csv-files/pubmedqa_rolling_recall.csv};
    \addlegendentry{\sys recall}
    \addplot+[color=Set1-E, solid, no marks, discard if not={series}{HNSW retrieval top-5 recall}]
      table[col sep=comma, x expr=\thisrow{event_index}/100000, y=value_pct]
      {csv-files/pubmedqa_rolling_recall.csv};
    \addlegendentry{HNSW recall}
    \addplot+[color=Set1-B, dashed, mark=*, discard if not={series}{ZeroRAG RAG accuracy}]
      table[col sep=comma, x expr=\thisrow{event_index}/100000, y=value_pct]
      {csv-files/pubmedqa_rolling_recall.csv};
    \addlegendentry{\sys RAG acc.}
    \addplot+[color=Set1-E, dashed, mark=*, discard if not={series}{HNSW RAG accuracy}]
      table[col sep=comma, x expr=\thisrow{event_index}/100000, y=value_pct]
      {csv-files/pubmedqa_rolling_recall.csv};
    \addlegendentry{HNSW RAG acc.}

    \nextgroupplot[
      xlabel={Queries issued ($\times 10^{5}$)},
      ylabel={Requests},
      ymin=0,
      legend columns=2,
    ]
    \addplot+[sys, name path=p10, opacity=0, forget plot] table[col sep=comma,
        x expr=\thisrow{window_end}/100000, y=p10]
      {csv-files/pubmedqa_zerorag_rolling_requests_to_found.csv};
    \addplot+[sys, name path=p90, opacity=0, forget plot] table[col sep=comma,
        x expr=\thisrow{window_end}/100000, y=p90]
      {csv-files/pubmedqa_zerorag_rolling_requests_to_found.csv};
    \addplot+[sys, opacity=0.15] fill between[of=p10 and p90];
    \addlegendentry{p10--p90}

    \addplot+[sys, name path=p25, opacity=0, forget plot] table[col sep=comma,
        x expr=\thisrow{window_end}/100000, y=p25]
      {csv-files/pubmedqa_zerorag_rolling_requests_to_found.csv};
    \addplot+[sys, name path=p75, opacity=0, forget plot] table[col sep=comma,
        x expr=\thisrow{window_end}/100000, y=p75]
      {csv-files/pubmedqa_zerorag_rolling_requests_to_found.csv};
    \addplot+[sys, opacity=0.35] fill between[of=p25 and p75];
    \addlegendentry{p25--p75}

    \addplot+[sys] table[col sep=comma,
        x expr=\thisrow{window_end}/100000, y=p50]
      {csv-files/pubmedqa_zerorag_rolling_requests_to_found.csv};
    \addlegendentry{median}

    \addplot+[sys-dashed] table[col sep=comma,
        x expr=\thisrow{window_end}/100000, y=p99]
      {csv-files/pubmedqa_zerorag_rolling_requests_to_found.csv};
    \addlegendentry{p99}
  \end{groupplot}
  \coordinate (capy) at ([yshift=-2pt]current bounding box.south);
  \node[anchor=north, font=\sffamily\small] at (g c1r1.south |- capy) {(a) Recall and RAG accuracy};
  \node[anchor=north, font=\sffamily\small, text width=3cm, align=center] at (g c2r1.south |- capy) {(b) Requests until the first relevant document is found};
\end{tikzpicture}
    \caption{(a) The \pubmedqa end-to-end \ac{RAG} performance
    , using \sys and \ac{HNSW}. 
    (b) the distribution of number of \sys requests 
    before
    the first relevant document.}
    \label{fig:pubmed_recall_requests}
\end{figure}

\subsection{End-to-End \ac{RAG} Performance}
\label{sec:expe2e}
We now compare the performance of document retrieval with \sys in an end-to-end \ac{RAG} pipeline, using the \pubmedqa workload.
In this experiment, peers retrieve documents using \sys and then use an \ac{LLM} (\texttt{Phi\-3.5-mini-instruct}~\cite{abdin2024phi}) to generate an answer.
We consider the generated answer to be successful if it matches the (yes, no, or maybe) ground truth answer provided by \pubmedqa. We use $k=5$ for document retrieval.

\Cref{fig:pubmed_recall_requests}(a) shows the document retrieval recall and \ac{RAG} accuracy for \sys and when using a centralized \ac{HNSW} index, which resembles an upper-bound on performance.
The solid and dashed lines represent recall and accuracy, respectively.
Both the recall and accuracy of \sys increase as peers issue searches, showing that \sys successfully adapts its routing to improve \ac{RAG} performance.
The recall and accuracy of the central \ac{HNSW} database remains constant across the experiment since it is a static data store.
At the end of the experiment, the \ac{HNSW} baseline reaches $97.3\%$ recall with an accuracy of $79\%$, while \sys reaches $93.0\%$ with $77\%$ accuracy.
The reason for the maximum recall of $97.3\%$, even with access to global data and the centralized index, is the imperfect embedding of queries and their (\textsc{GPT5-nano}) rephrasings: it could be that an embedding of a document relevant to a query is not in the top-5 closest embeddings.
This 
shows that \sys gets performance relatively close to that of a centralized baseline.

\sys can act as a progressive retrieval mechanism: a peer may contact only a few other peers for a fast, approximate result, or contact more peers to improve retrieval quality at the cost of additional messages.
We next measure how many request messages (see \Cref{sec:zerorag-design}) are needed until \sys finds the closest document that exists in the network as the overlay organizes.
\Cref{fig:pubmed_recall_requests}(b) shows this information, grouped by percentile range and including the median.
We observe that as the overlay self-organizes over time, the number of query messages required to find the best document decreases.
In the final state
, 99\% of queries require no more than \num{13} query messages to find the best supporting document.

On average, the centralized \ac{HNSW} baseline processes its longest-duration queries in $24$~ms\slash query, whereas \sys requires $379$~ms\slash query (excluding network latency).
This difference is the expected cost of decentralization: a centralized index can exploit global knowledge and optimized single-machine data structures, whereas \sys achieves comparable recall through purely local interactions across an overlay of peers, without any central authority, global index, or centralized data stores.

\colorlet{shadecolor}{takeawaycolor}
\begin{shaded}
\noindent\textbf{Takeaway (\Cref{fig:pubmed_recall_requests}).} End-to-end \ac{RAG} performance with \sys approaches the recall and accuracy of the centralized \ac{HNSW} baseline.
For $99\%$ of queries, \sys can find the best document in a network in just $13$ requests.
\end{shaded}

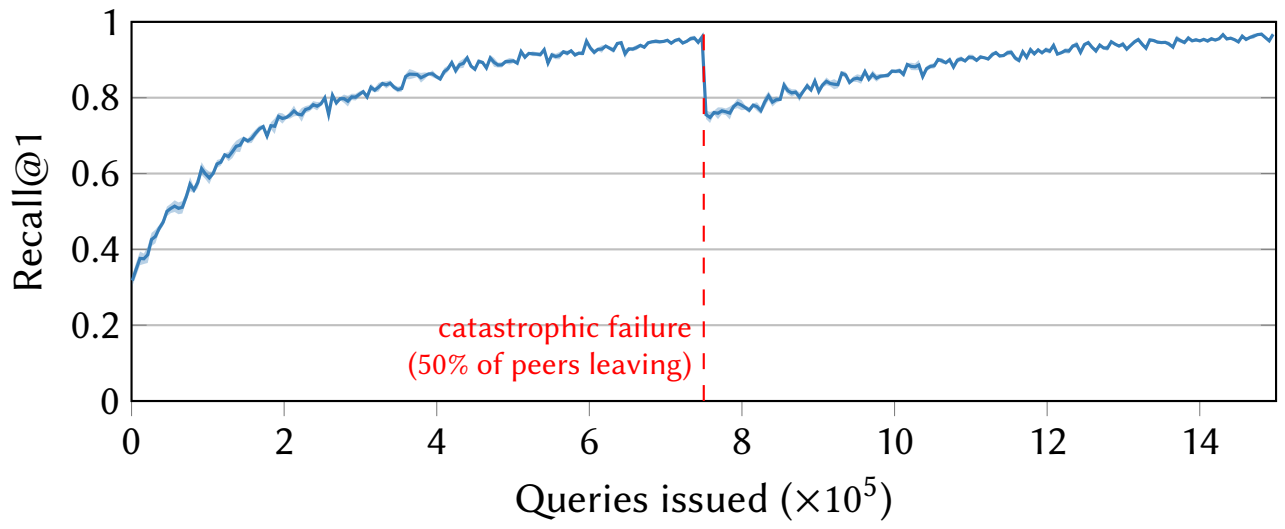
\begin{figure}[t]
    \centering
\begin{tikzpicture}
  \begin{axis}[
    xlabel={Queries issued ($\times 10^{5}$)},
    ylabel={Recall@1},
    ymin=0, ymax=1,
    xmin=0, xmax=15,
    width=260.4pt,
    scaled x ticks=false,
    xticklabel style={/pgf/number format/.cd, fixed, precision=0},
    every axis plot/.append style={
      very thick, no marks,
      each nth point=50, unbounded coords=jump,
    },
  ]
    \addplot[name path=lo, draw=none, forget plot] table[col sep=comma,
        x expr=\thisrow{event_index}/100000, y=recall_ci95_lower]
      {csv-files/churn_rolling_recall_1000_events_drop50.csv};
    \addplot[name path=hi, draw=none, forget plot] table[col sep=comma,
        x expr=\thisrow{event_index}/100000, y=recall_ci95_upper]
      {csv-files/churn_rolling_recall_1000_events_drop50.csv};
    \addplot[Set1-B, fill opacity=0.35, draw=none, forget plot] fill between[of=hi and lo];
    \addplot+[sys, semithick] table[col sep=comma,
        x expr=\thisrow{event_index}/100000, y=recall]
      {csv-files/churn_rolling_recall_1000_events_drop50.csv};
    \draw[dashed, red] (axis cs:7.49952,0) -- (axis cs:7.49952,1);
    \node[anchor=south east, font=\sffamily\scriptsize, red, align=right]
      at (axis cs:7.49952,0.02) {catastrophic failure\\(50\% of peers leaving)};
  \end{axis}
\end{tikzpicture}
    \caption{Average recall@1 per processed query batch in \sys with the \pubmedqa workload, when there is a catastrophic failure event (dashed red line).}
    \label{fig:churn}
\end{figure}

\subsection{Resilience to Catastrophic Peer Failure}
\label{sec:expchurn}
We now analyze \sys's resilience under a catastrophic failure event, in which 50\% of all peers go offline simultaneously at the midpoint of the experiment and are gradually replaced by new peers over the remainder of the run.
For this experiment, we use the \pubmedqa workload and we compute the average recall@1 for each batch of \num{1000} queries performed.
\Cref{fig:churn} shows these results and we annotate the catastrophic failure event with a vertical dashed line.
Recall drops sharply at the moment of failure because the learned paths and routed-query memories of the departed peers are lost.
Nevertheless, \sys recovers to 95.5\% recall in the final 10\% of the run.
This recovery happens entirely through local interactions: replacement peers accumulate routed-query memory and adaptive edges as they receive search queries, rebuilding useful routing state.
After the catastrophic failure, peers require 15.1 requests per query to find the relevant document, which is higher compared to the same number before the catastrophic event (8.9):
the neighborhoods of newly-joined peers are not yet optimal.

\colorlet{shadecolor}{takeawaycolor}
\begin{shaded}
\noindent\textbf{Takeaway (\Cref{fig:churn}).} \sys is robust against catastrophic failure where routing dynamics get disrupted significantly, and self-heals entirely through local interactions as replacement peers rebuild routing state from incoming queries.
\end{shaded}

\subsection{Ablation Study}
\label{sec:ablation}
We 
analyze the effect of key parameters and components on 
routing performance.
Unless stated otherwise, experiments use the \pubmedqa workload and results are reported over the final 10\% of queries, by which point the overlay has converged.
We report 
\emph{average recall@1}, measuring the fraction of target documents successfully retrieved, and \emph{average requests required}, measuring the number of \textsc{query} messages sent until the target document is found.
Together, these capture both 
routing
effectiveness and efficiency. 


\begin{figure}[t]
    \centering
\definecolor{zsblue}{HTML}{4878A8}      
\definecolor{zslightblue}{HTML}{9EC9E2} 
\definecolor{zsnavy}{HTML}{2B3D63}      
\definecolor{zspurple}{HTML}{7D4F9C}    
\definecolor{ablfull}{HTML}{0072B2}     
\definecolor{ablnomem}{HTML}{D55E00}    
\definecolor{ablstatic}{HTML}{009E73}   

\begin{tikzpicture}
  \begin{groupplot}[
    group style={
      group size=2 by 1,
      horizontal sep=1.3cm,
      group name=g,
    },
    width=134.34pt,
    legend style={font=\sffamily\scriptsize},
  ]
    \nextgroupplot[
      xlabel={Query batches processed},
      ylabel={Average recall@1},
      ymin=0.7, ymax=1,
      xmin=0,
      legend columns=1,
      legend style={font=\sffamily\scriptsize, at={(0.5,1.02)}, anchor=south},
      every axis plot/.append style={semithick, no marks, solid, smooth, each nth point=20},
    ]
    \addplot[color=ablfull] table[col sep=comma, x=window_index, y={ZeroRAG}]
      {csv-files/aol_recall_rolling1000.csv};
    \addlegendentry{\sys}
    \addplot[color=ablnomem] table[col sep=comma, x=window_index, y={ZeroRAG no query memory}]
      {csv-files/aol_recall_rolling1000.csv};
    \addlegendentry{\sys (no query memory)}
    \addplot[color=ablstatic] table[col sep=comma, x=window_index, y={ZeroRAG no new connections}]
      {csv-files/aol_recall_rolling1000.csv};
    \addlegendentry{\sys (static topology)}

    \nextgroupplot[
      xlabel={Available relevant docs per query},
      xlabel style={text width=3cm, align=center},
      ylabel={Retrieved relevant docs (\%)},
      ymin=0, ymax=100,
      ybar=0.5pt,
      bar width=4.5pt,
      symbolic x coords={1-10, 11-50, 51-100},
      xtick=data,
      enlarge x limits=0.25,
      bar legend,
      legend columns=1,
      legend style={font=\sffamily\scriptsize, at={(0.5,1.02)}, anchor=south,
        /tikz/every even column/.append style={column sep=4pt}},
      every axis plot/.append style={fill, draw=none},
    ]
    \addplot[std bars, draw=none, fill=zsblue]      table[col sep=comma, x=relevant_docs_bin, y expr=\thisrow{zerorag_mean_query_retrieval_recall}*100,
        y error expr=\thisrow{zerorag_mean_query_retrieval_recall_std}*100]
      {csv-files/amazon_relevant_retrieval_histogram.csv};
    \addlegendentry{\sys}
    \addplot[std bars, draw=none, fill=zslightblue] table[col sep=comma, x=relevant_docs_bin, y expr=\thisrow{zerorag_no_query_memory_mean_query_retrieval_recall}*100,
        y error expr=\thisrow{zerorag_no_query_memory_mean_query_retrieval_recall_std}*100]
      {csv-files/amazon_relevant_retrieval_histogram.csv};
    \addlegendentry{\sys (no query memory)}
    \addplot[std bars, draw=none, fill=zsnavy]      table[col sep=comma, x=relevant_docs_bin, y expr=\thisrow{zerorag_no_new_connections_mean_query_retrieval_recall}*100,
        y error expr=\thisrow{zerorag_no_new_connections_mean_query_retrieval_recall_std}*100]
      {csv-files/amazon_relevant_retrieval_histogram.csv};
    \addlegendentry{\sys (static topology)}
    \addplot[std bars, draw=none, fill=zspurple]    table[col sep=comma, x=relevant_docs_bin, y expr=\thisrow{zerorag_no_query_memory_no_new_connections_mean_query_retrieval_recall}*100,
        y error expr=\thisrow{zerorag_no_query_memory_no_new_connections_mean_query_retrieval_recall_std}*100]
      {csv-files/amazon_relevant_retrieval_histogram.csv};
    \addlegendentry{\sys (no mem, static)}
  \end{groupplot}
  \coordinate (capy) at ([yshift=-2pt]current bounding box.south);
  \node[anchor=north, font=\sffamily\small] at (g c1r1.south |- capy) {(a) \aolia};
  \node[anchor=north, font=\sffamily\small] at (g c2r1.south |- capy) {(b) \amazon};
\end{tikzpicture}
    \caption{Ablation of query memory and adaptive topology on the (a) \aolia and (b) \amazon workloads. 
    }
    \label{fig:ablation-inputs}
\end{figure}
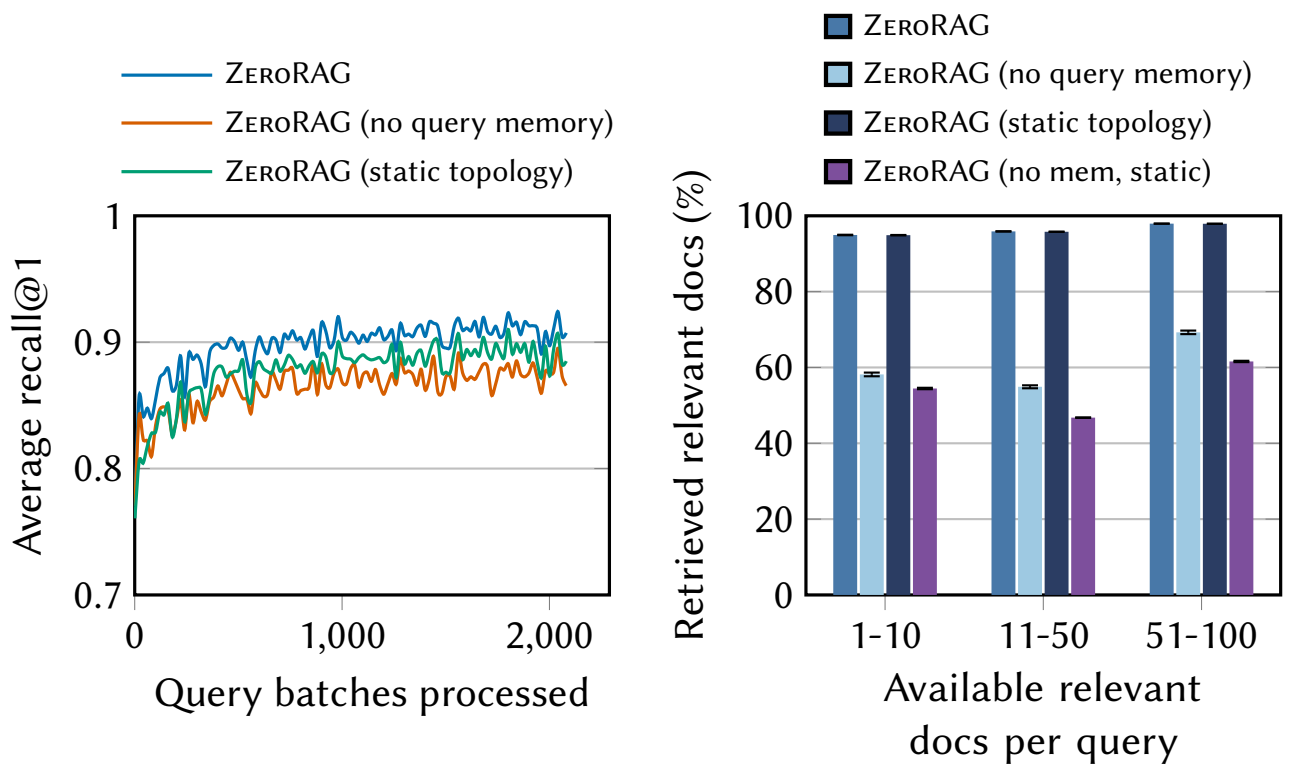

\subsubsection{The effect of query cache and topology adaptation.}
\Cref{fig:ablation-inputs} isolates the effect of \sys's two routing inputs by disabling routed-query memory, adaptive topology formation, or both, and comparing against the full system on both the \aolia and \amazon workloads.
Routed-query memory is the dominant contributor on both workloads: on \aolia its removal lowers average recall@1 from \num{0.89} to \num{0.85}, while on the more demanding multi-relevant \amazon coverage task it has a much larger effect.
Disabling topology adaptation alone leaves retrieval close to the full system, and removing both inputs degrades it the most, confirming that the two mechanisms are complementary.

\begin{table}[t]
\centering
\small
\caption{The effect of query cache size.}
\label{tab:pubmedqa-query-memory-limit}
\begin{tabular}{lrr}
\toprule
Max. cache size ($\delta_m$) & Avg. Recall@1 & Avg. requests required \\
\midrule
0 & $(77.860 \pm 0.177)\%$ & 16.796 $\pm$ 0.067 \\
10 & $(85.834 \pm 0.120)\%$ & 12.757 $\pm$ 0.041 \\
100 & $(97.612 \pm 0.037)\%$ & 5.900 $\pm$ 0.017 \\
1,000 & $(99.652 \pm 0.007)\%$ & 3.448 $\pm$ 0.006 \\
10,000 & $(99.652 \pm 0.007)\%$ & 3.448 $\pm$ 0.006 \\
\bottomrule
\end{tabular}
\end{table}

\subsubsection{The effect of query routing cache size ($\delta_m$).}
We vary the size of the query routing cache ($\delta_m$) and report the average recall@1 and number of requests required to obtain a document for the last 10\% of the queries (of the \pubmedqa workload).
\Cref{tab:pubmedqa-query-memory-limit} shows that $\delta_m$ is a key contributor to the routing effectiveness of \sys.
With no routed-query memory ($\delta_m=0$), \sys obtains 77.9\% recall and requires 16.8 requests
to find the target document.
Storing 10 routed queries ($\delta_m=10$) 
improves recall to 85.8\% and reduces the average required requests to 12.75.
Increasing $\delta_m$ 
to 100 improves recall to 97.6\% and the request count drops to 5.9.
Even higher values 
lead to 
diminishing returns for recall and required requests.



\begin{table}[t]
\centering
\small
\caption{The effect of network size.}
\label{tab:pubmedqa-user-scalability}
\begin{tabular}{lrr}
\toprule
Network size (peers) & Avg. Recall@1 & Avg. requests required \\
\midrule
1,000 & $(99.365 \pm 0.272)\%$ & 3.209 $\pm$ 0.111 \\
5,000 & $(98.276 \pm 0.066)\%$ & 4.330 $\pm$ 0.059 \\
10,000 & $(98.378 \pm 0.051)\%$ & 4.656 $\pm$ 0.015 \\
50,000 & $(99.652 \pm 0.007)\%$ & 3.448 $\pm$ 0.006 \\
100,000 & $(99.913 \pm 0.005)\%$ & 5.724 $\pm$ 0.011 \\
\bottomrule
\end{tabular}
\end{table}

\subsubsection{Scalability.}
We evaluate the routing effectiveness and efficiency as the number of peers in the network increases.
We vary the network size from \num{1000} to \num{100000} peers and analyze the average recall@1 and number of requests required to obtain a document for the last 10\% of the queries. 
Because of our setup, there could be a situation where a peer attempts to retrieve a document that is not advertised by any other peer in the network.
We exclude these queries when computing recall@1, so recall is measured conditional on the target document being advertised by at least one remote peer.
\Cref{tab:pubmedqa-user-scalability} shows that average recall@1 remains high (above 98\%), even as the network size grows to \num{100000} peers.
This indicates that, once a relevant document exists in the network, \sys can locate it with high probability even as the peer population grows by two orders of magnitude.
The average request count is non-monotonic as the network size increases.
Furthermore, the average number of requests required to obtain the target document remains low.

\begin{table}[t]
\centering
\small
\caption{The effect of document count per peer.}
\label{tab:pubmedqa-doc-scalability}
\begin{tabular}{lrr}
\toprule
Docs. per peer ($t$) & Avg. Recall@1 & Avg. requests required \\
\midrule
10 & $(98.665 \pm 0.018)\%$ & 5.940 $\pm$ 0.023 \\
20 & $(99.292 \pm 0.016)\%$ & 5.545 $\pm$ 0.017 \\
30 & $(99.667 \pm 0.007)\%$ & 5.361 $\pm$ 0.012 \\
40 & $(99.836 \pm 0.006)\%$ & 5.276 $\pm$ 0.009 \\
50 & $(99.919 \pm 0.004)\%$ & 5.222 $\pm$ 0.007 \\
\bottomrule
\end{tabular}
\end{table}

\subsubsection{Documents per peer.}
We evaluate the routing effectiveness and efficiency as we assign $t$ more \pubmedqa documents to individual peers. 
We show average recall@1 and requests required in \Cref{tab:pubmedqa-doc-scalability}.
Even for $t=10$, recall 
reaches 98.66\% and increases to 99.91\% for $t=50$.
As documents 
are
more available, 
the number of requests 
to obtain them decreases from 5.94 for $t=10$ to 5.22 for $t=50$: a decrease of 11.5\%.
We conclude that \sys is effective even with small local document stores, while larger stores substantially improve coverage and reduce message overhead.


\begin{table}[t]
\centering
\small
\caption{The effect of neighborhood size. Five seeds.}
\label{tab:pubmedqa-degree}
\begin{tabular}{lrr}
\toprule
Max. neighbors ($\delta_n$) & Avg. Recall@1 & Avg. requests required \\
\midrule
3 & $(98.614 \pm 0.022)\%$ & 6.198 $\pm$ 0.020 \\
5 & $(99.080 \pm 0.012)\%$ & 5.360 $\pm$ 0.009 \\
10 & $(99.417 \pm 0.012)\%$ & 4.431 $\pm$ 0.010 \\
25 & $(99.606 \pm 0.004)\%$ & 3.702 $\pm$ 0.010 \\
50 & $(99.644 \pm 0.010)\%$ & 3.505 $\pm$ 0.005 \\
75 & $(99.666 \pm 0.008)\%$ & 3.388 $\pm$ 0.004 \\
100 & $(99.675 \pm 0.012)\%$ & 3.305 $\pm$ 0.006 \\
\bottomrule
\end{tabular}
\end{table}

\subsubsection{Neighborhood size ($\delta_n$).}
We evaluate routing effectiveness and efficiency as we vary the number of neighbors ($\delta_n$) each peer is connected to, see \Cref{tab:pubmedqa-degree}.
Varying $\delta_n$ from 3 to \num{100} has a moderate effect on recall: 98.61\% for $\delta_n=3$ compared to 99.7\% for $\delta_n=100$.
Increasing $\delta_n$ shows diminishing returns: 
a small neighborhood size is sufficient to obtain high recall as the network converges.
In contrast, the average number of 
requests 
shows a sharper decrease: 6.2 for $\delta_n=3$ to 3.30 for $\delta_n=100$, a decrease of 46.5\%.
Larger values of $\delta_n$ primarily reduce the number of requests 
by giving each peer more immediate semantic routing options.
This result suggests that moderate neighborhood sizes (\eg, 50) are reasonable default choices, striking a good balance between recall and efficiency.


\begin{table}[t]
\centering
\small
\caption{The effect of parallel searches.}
\label{tab:pubmedqa-fanout}
\begin{tabular}{llrr}
\toprule
Fanout & Rounds & Avg. Recall@1 & Avg. rounds required \\
\midrule
3 & 10 & $(99.374 \pm 0.010)\%$ & 1.718 $\pm$ 0.002 \\
6 & 5 & $(99.203 \pm 0.012)\%$ & 1.508 $\pm$ 0.001 \\
10 & 3 & $(97.456 \pm 0.040)\%$ & 1.408 $\pm$ 0.001 \\
\bottomrule
\end{tabular}
\end{table}

\subsubsection{Parallel requests.}
We evaluate 
how
parallelizing search 
affects 
recall and the number of rounds 
to find a target document.
A peer sends $f$ \textsc{query} and \textsc{queryResult} in parallel for a fixed number of rounds.
We fix the contact budget $C=30$ and redistribute it across fanout and number of rounds: 3 requests per round for up to 10 rounds, 6 requests per round for 5 rounds, and 10 requests per round for 3 rounds.
\Cref{tab:pubmedqa-fanout} reports the  recall and average rounds 
over the final 10\% of queries.
The $f=3$ configuration reaches 99.37\% recall and requires 1.72 rounds to find the target document, 
while increasing $f$ to 6 we get almost the same recall, 99.20\%.
With $f=10$, recall drops very slightly to 97.46\%.
A query typically requires several referral steps to reach the right semantic neighborhood.
Consequently, parallelization is most beneficial in low-traffic networks where the added message volume is acceptable, as it reduces wall-clock latency without sacrificing routing depth.

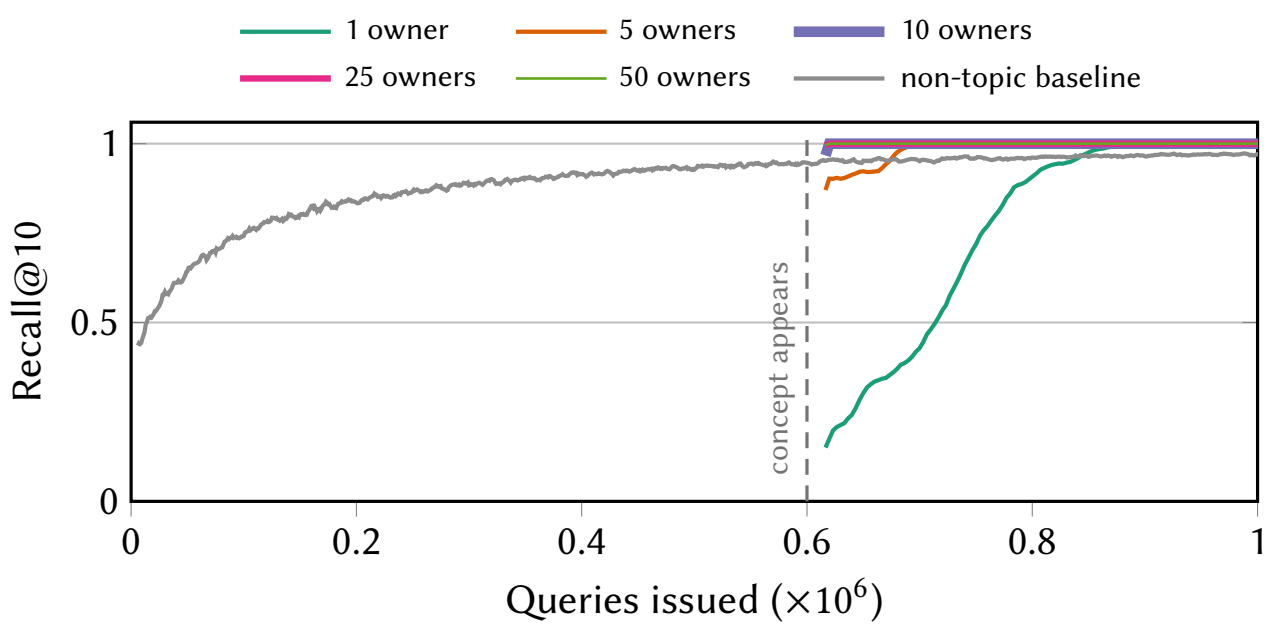
\begin{figure}[t]
    \centering
%
\begin{tikzpicture}
  \begin{axis}[
    xlabel={Queries issued ($\times 10^{6}$)},
    ylabel={Recall@10},
    ymin=0, ymax=1.06,
    xmin=0, xmax=1.0,
    width=257pt,
    cycle list name=Dark2,
    legend style={font=\sffamily\scriptsize, at={(0.5,1.02)}, anchor=south,
      /tikz/every even column/.append style={column sep=5pt}},
    legend columns=3,
    every axis plot/.append style={thick, no marks, solid},
    unbounded coords=discard,
  ]
    \addplot+[discard if not={series}{1 owners topic}]
      table[col sep=comma, x expr=\thisrow{event_index}/1000000, y=rolling_recall]
      {csv-files/topic_addition_owner_topic_overlay.csv};
    \addlegendentry{1 owner}
    \addplot+[discard if not={series}{5 owners topic}]
      table[col sep=comma, x expr=\thisrow{event_index}/1000000, y=rolling_recall]
      {csv-files/topic_addition_owner_topic_overlay.csv};
    \addlegendentry{5 owners}
    \addplot+[line width=2.0pt, discard if not={series}{10 owners topic}]
      table[col sep=comma, x expr=\thisrow{event_index}/1000000, y=rolling_recall]
      {csv-files/topic_addition_owner_topic_overlay.csv};
    \addlegendentry{10 owners}
    \addplot+[line width=1.2pt, discard if not={series}{25 owners topic}]
      table[col sep=comma, x expr=\thisrow{event_index}/1000000, y=rolling_recall]
      {csv-files/topic_addition_owner_topic_overlay.csv};
    \addlegendentry{25 owners}
    \addplot+[line width=0.5pt, discard if not={series}{50 owners topic}]
      table[col sep=comma, x expr=\thisrow{event_index}/1000000, y=rolling_recall]
      {csv-files/topic_addition_owner_topic_overlay.csv};
    \addlegendentry{50 owners}
    \addplot[draw=black!45, discard if not={series}{non-topic baseline (1 owner run)}]
      table[col sep=comma, x expr=\thisrow{event_index}/1000000, y=rolling_recall]
      {csv-files/topic_addition_owner_topic_overlay.csv};
    \addlegendentry{non-topic baseline}

    \draw[densely dashed, black!55] (axis cs:0.6,0) -- (axis cs:0.6,1.02);
    \node[rotate=90, anchor=south west, font=\sffamily\scriptsize, black!55]
      at (axis cs:0.6,0.04) {concept appears};
  \end{axis}
\end{tikzpicture}
    \caption{
    Introduction of a new topic at the 60\% mark (vertical line).
    The lines of owners $\ge 10$ overlap.
    }
    \label{fig:concept_drift}
\end{figure}

\subsection{Spreading New Concepts}
\label{sec:expdrift}
Our last experiment shows
how fast \sys spreads a new topic in a network, analyzing how recall evolves for topics which appear anew, and are very sparsely available, but suddenly become popular with all peers.
To do so, we omit a single topic from the \pubmedqa workload 
and re-introduce it at the 60\% mark into the experiment.
We chose a random topic (``Peptostreptococcus'') that has 10 relevant documents, and we vary how many peers own it initially.
All peers then continuously send queries to find documents on our chosen (initially omitted) topic.
A document starts appearing in other peers' local data stores only once a query successfully reaches it. 

\Cref{fig:concept_drift} shows the rolling recall@10 for both the topic's queries per initial owner count and the average for other topics as a gray line.
Seeding the topic at more owners lets recall recover faster once the concept reappears: with 10 or more owners the documents are immediately reachable across the network and the recall becomes $1.0$ almost instantly, causing the lines to overlap.
With less initial owners the documents also spread less fast, with the single-owner case being the slowest, climbing back only as the documents propagate through the overlay once queries begin to reach them. 

\colorlet{shadecolor}{takeawaycolor}
\begin{shaded}
\noindent\textbf{Takeaway (\Cref{fig:concept_drift}).} \sys adapts to newly introduced, initially sparse topics through local interactions alone, with its speed governed by how widely the new content is initially held.
\end{shaded}
\section{Related Work}

We discuss works on
\ac{RAG} and 
nearest-neighbor search, 
federated and distributed \ac{RAG} that scale retrieval without removing a coordinating authority, and decentralized peer-to-peer retrieval and routing, which is closest to our setting.
\Cref{tab:system_comparison} gives an overview.

\textbf{Retrieval-augmented generation.}
\Ac{RAG} grounds \ac{LLM} generation in external documents retrieved at query time, improving factual accuracy without retraining~\cite{lewis2020retrieval,fan2024survey,zhao2026retrieval}.
Retrieval is typically cast as nearest-neighbor search over high-dimensional embeddings: queries and documents are embedded into a shared space, and the documents closest to the query are returned~\cite{nene1997simple}.
Because exact search is expensive at scale, practical systems rely on \ac{ANN} indexes such as \ac{HNSW} graphs~\cite{malkov2018efficient} or \ac{LSH}~\cite{LSH-pstable-Dist,li2019approximate}.
These techniques assume documents are collected and indexed under a single administrative domain, so the index can be built globally.
\sys targets the opposite: when documents remain with their owners and no global index can be constructed.

\begin{sloppypar}
\textbf{Federated and distributed RAG.}
Federated approaches route or merge results across independently managed collections: RAGRoute selects which remote sources to query for a given request~\cite{RAGRoute}, FeB4RAG benchmarks federated search for \ac{RAG}~\cite{FeB4RAG}, and pFedRAG personalizes retrieval models across federated clients~\cite{he-etal-2025-pfedrag}.
Others target the privacy or efficiency of the retrieval
: FedVSE provides privacy-preserving vector search over federated databases~\cite{FedVSE}, PRAG offers end-to-end privacy-preserving \ac{RAG}~\cite{li2026pragendtoendprivacypreservingretrievalaugmented}, and Proximity caches semantically similar queries to avoid repeated vector-database lookups~\cite{Proximity}.
A parallel line scales the retrieval substrate itself: DRAGON distributes \ac{RAG} retrieval across local and cloud resources~\cite{dragon}, DGRAG organizes graph-based retrieval over edge-cloud systems~\cite{zhou2026dgragdistributedgraphbasedretrievalaugmented}, and HARMONY and SPIRE build scalable distributed vector databases for high-throughput \ac{ANN} search~\cite{Harmony,spire}.
These systems improve scalability, privacy, or cost, but assume a federated, cloud, or managed deployment in which a coordinating component still orchestrates retrieval.
\sys instead routes each query through peer-local state alone. 
\end{sloppypar}

\textbf{Decentralized retrieval and routing.}
The closest systems to \sys 
use
a decentralized overlay 
and a local or global state.
One family imposes a shared logical structure that must be maintained as peers and documents change.
Semantica organizes peers into a global \ac{LLM}-guided semantic tree overlay~\cite{semantica}, while \ac{DHT} designs such as Kademlia and \ac{LSH}-based distributed indexes place keys in a global address space for approximate lookup~\cite{kademlia,LSH-pstable-Dist}.
Blockchain-based designs decentralize control through shared ledger state: D-RAG coordinates decentralized \ac{RAG} through a blockchain~\cite{d-rag}, and a related system records a reliability score for each document source on a blockchain so that the trustworthiness of retrieved content cannot be tampered with~\cite{lu2025decentralizedretrievalaugmentedgeneration}.
These approaches remain decentralized but trade a central authority for global coordination.

A second family keeps routing state local but depends on explicit symbolic descriptions.
REMINDIN' routes semantic queries using social metaphors and observations of previously successful peers~\cite{Remindin}; DRAG/TARW performs topic-aware random walks given global knowledge of the available topics~\cite{drag/tarw}; and SHIMI organizes decentralized agent memory through an explicit semantic hierarchy~\cite{helmi2025decentralizingaimemoryshimi}.
These systems share \sys's local-state philosophy, but rely on topic labels, expertise categories, or maintained hierarchies rather than routing directly from unlabeled embeddings.

A third, longer line of decentralized recommendation and nearest-neighbor work anticipates our overlay mechanics.
WHATSUP disseminates news recommendations through an epidemic protocol that clusters users by interest without any central server~\cite{6569858}, and a follow-up obfuscation-based protocol makes such profile-based collaborative filtering privacy-preserving~\cite{Boutet2016}.
Most directly, Cluster-and-Conquer constructs a $k$-nearest-neighbor graph by starting from a random overlay and greedily improving it through neighbor-of-neighbor exploration~\cite{9458649}, the same locality-exploiting principle \sys uses to rewire its overlay.
These build on classical decentralized overlay techniques such as gossip-based peer sampling~\cite{jelasity2007gossip}, semantic-proximity content search~\cite{voulgaris2004exploiting}, and the broader literature on robust peer-to-peer search~\cite{risson2007survey}.
\sys routes from unlabeled query and document embeddings and, crucially, additionally remembers recently routed queries, letting peers steer toward neighbors with semantically similar request histories. 

Overall, prior work occupies only part of the design space.
Shared-structure methods are decentralized and often unlabeled, but require global maintenance; some local routing methods reduce coordination, but rely on labeled or 
structured semantics; and federated vector-search systems improve scalability, but are not fully decentralized.
To our knowledge, \sys is unique in treating retrieval as a querier-based local routing problem: it avoids the global maintenance that \semantica and Kademlia require, while remaining fully decentralized, unlike centralized or federated alternatives.

\begin{table}
  \caption{Whether various retrieval and routing systems have (``Decentr.'') decentralized governance,  use (``Unlab.'') unlabeled data, or use a global, local, or no (``N/A'') data structure.
  }
  \label{tab:system_comparison}
  \small
  \begin{tabular}{p{3.4cm}ccc}
    \toprule
    System&Decentr.&Unlab.&Data Structure\\
    \midrule
        \textbf{\sys [this work]} & \checkmark & \checkmark & \textbf{Local} \\
        \semantica~\cite{semantica} & \checkmark & \checkmark & Global \\
        \textsc{LSH}/\textsc{Kademlia}~\cite{LSH-pstable-Dist, kademlia} & \checkmark & \checkmark & Global \\
        \textsc{REMINDIN'}~\cite{Remindin} & \checkmark &  & Local \\
        \textsc{SHIMI}~\cite{helmi2025decentralizingaimemoryshimi} & \checkmark &  & Local \\
        \textsc{Proximity}~\cite{Proximity} & & \checkmark & N/A \\
        \textsc{DRAG} (\textsc{TARW})~\cite{drag/tarw} & \checkmark & & Local \\
        \textsc{pFedRAG}~\cite{he-etal-2025-pfedrag} &  & \checkmark & N/A \\
        \textsc{DRAGON}~\cite{dragon} & & \checkmark & N/A \\
        \textsc{D-RAG} (Blockchain)~\cite{d-rag} & \checkmark & & Global \\
        \textsc{HARMONY}/\textsc{SPIRE}~\cite{Harmony,spire} &  & \checkmark & N/A \\
\bottomrule
\end{tabular}
\end{table}
\section{Summary}

We presented \sys, a decentralized routing layer for \ac{RAG} that locates relevant documents across an overlay of peers without any central authority, labeled data, or global index.
Each peer keeps only a bounded set of neighbor advertisements and a small cache of recently routed queries, routing each query through a best-first walk that continuously reconfigures the overlay toward semantically relevant neighbors.
Our evaluation on the \aolia, \amazon, and \pubmedqa workloads shows that \sys consistently outperforms \semantica and \dhtlsh, approaches the recall of a centralized \ac{HNSW} index in an end-to-end \ac{RAG} pipeline ($93.0\%$ vs.\ $97.3\%$ at comparable accuracy).
\sys 
recovers to $95.5\%$ recall after losing half the network, scales to \num{100000} peers, and absorbs new topics through local interactions alone.
These results indicate that decentralization is a practical foundation for \ac{RAG}, even when documents 
remain with the peers that own them.

\section*{Availability}
The implementation of this work is available at \url{https://github.com/pneague/ZeroRAG-System-Simulation}.

\begin{acks}
This work has been funded by the Swiss National Science Foundation, under the project "FRIDAY: Frugal, Privacy-Aware and Practical Decentralized Learning", SNSF grant number 10001796, and by NWO/TKI grant BLOCK.2019.004
\end{acks}

\bibliographystyle{ACM-Reference-Format}
\bibliography{references}

@misc{li2026pragendtoendprivacypreservingretrievalaugmented,
      title={PRAG: End-to-End Privacy-Preserving Retrieval-Augmented Generation}, 
      author={Zhijun Li and Minghui Xu and Huayi Qi and Wenxuan Yu and Tingchuang Zhang and Qiao Zhang and GuangYong Shang and Zhen Ma and Xiuzhen Cheng},
      year={2026},
      eprint={2604.26525},
      archivePrefix={arXiv},
      primaryClass={cs.CR},
}

@misc{spire,
      title={Scalable Distributed Vector Search via Accuracy Preserving Index Construction}, 
      author={Yuming Xu and Qianxi Zhang and Qi Chen and Baotong Lu and Menghao Li and Philip Adams and Mingqin Li and Zengzhong Li and Jing Liu and Cheng Li and Fan Yang},
      year={2025},
      eprint={2512.17264},
      archivePrefix={arXiv},
      primaryClass={cs.DC},
}

@inproceedings{Remindin,
author = {Tempich, Christoph and Staab, Steffen and Wranik, Adrian},
title = {Remindin': semantic query routing in peer-to-peer networks based on social metaphors},
year = {2004},
isbn = {158113844X},
publisher = {Association for Computing Machinery},
address = {New York, NY, USA},
url = {https://doi.org/10.1145/988672.988759},
doi = {10.1145/988672.988759},
booktitle = {Proceedings of the 13th International Conference on World Wide Web},
pages = {640–649},
numpages = {10},
location = {New York, NY, USA},
series = {WWW '04}
}

@inproceedings{RAGRoute, author = {Guerraoui, Rachid and Kermarrec, Anne-Marie and Petrescu, Diana and Pires, Rafael and Randl, Mathis and de Vos, Martijn},
title = {Efficient Federated Search for Retrieval-Augmented Generation},
year = {2025},
isbn = {9798400715389},
publisher = {Association for Computing Machinery},
address = {New York, NY, USA},
url = {https://doi.org/10.1145/3721146.3721942},
doi = {10.1145/3721146.3721942},
booktitle = {Proceedings of the 5th Workshop on Machine Learning and Systems},
pages = {74–81},
numpages = {8},
location = {World Trade Center, Rotterdam, Netherlands},
series = {EuroMLSys '25} }

@misc{zhou2026dgragdistributedgraphbasedretrievalaugmented,
      title={DGRAG: Distributed Graph-based Retrieval-Augmented Generation in Edge-Cloud Systems}, 
      author={Wenqing Zhou and Yuxuan Yan and Qianqian Yang},
      year={2026},
      eprint={2505.19847},
      archivePrefix={arXiv},
      primaryClass={cs.AI},
}

@inproceedings{he-etal-2025-pfedrag,
  title = "p{F}ed{RAG}: A Personalized Federated Retrieval-Augmented Generation System with Depth-Adaptive Tiered Embedding Tuning",
    author = "He, Hangyu  and
      Yuan, Xin  and
      Wu, Kai  and
      Liu, Ren Ping  and
      Ni, Wei",
    editor = "Christodoulopoulos, Christos  and
      Chakraborty, Tanmoy  and
      Rose, Carolyn  and
      Peng, Violet",
    booktitle = "Findings of the Association for Computational Linguistics: EMNLP 2025",
    month = nov,
    year = "2025",
    address = "Suzhou, China",
    publisher = "Association for Computational Linguistics",
    url = "https://aclanthology.org/2025.findings-emnlp.769/",
    doi = "10.18653/v1/2025.findings-emnlp.769",
    pages = "14255--14268",
    ISBN = "979-8-89176-335-7"
}

@inproceedings{FeB4RAG,
author = {Wang, Shuai and Khramtsova, Ekaterina and Zhuang, Shengyao and Zuccon, Guido},
title = {FeB4RAG: Evaluating Federated Search in the Context of Retrieval Augmented Generation},
year = {2024},
isbn = {9798400704314},
publisher = {Association for Computing Machinery},
address = {New York, NY, USA},
url = {https://doi.org/10.1145/3626772.3657853},
doi = {10.1145/3626772.3657853},
booktitle = {Proceedings of the 47th International ACM SIGIR Conference on Research and Development in Information Retrieval},
pages = {763–773},
numpages = {11},
location = {Washington DC, USA},
series = {SIGIR '24}
}

@article{Harmony,
author = {Xu, Qian and Zhang, Feng and Li, Chengxi and Cao, Lei and Chen, Zheng and Zhai, Jidong and Du, Xiaoyong},
title = {HARMONY: A Scalable Distributed Vector Database for High-Throughput Approximate Nearest Neighbor Search},
year = {2025},
issue_date = {September 2025},
publisher = {Association for Computing Machinery},
address = {New York, NY, USA},
volume = {3},
number = {4},
url = {https://doi.org/10.1145/3749167},
doi = {10.1145/3749167},
journal = {Proc. ACM Manag. Data},
month = sep,
articleno = {249},
numpages = {28}
}

@article{malkov2018efficient,
author = {Malkov, Yu A. and Yashunin, D. A.},
title = {Efficient and Robust Approximate Nearest Neighbor Search Using Hierarchical Navigable Small World Graphs},
year = {2020},
issue_date = {April 2020},
publisher = {IEEE Computer Society},
address = {USA},
volume = {42},
number = {4},
issn = {0162-8828},
url = {https://doi.org/10.1109/TPAMI.2018.2889473},
doi = {10.1109/TPAMI.2018.2889473},
journal = {IEEE Trans. Pattern Anal. Mach. Intell.},
month = apr,
pages = {824–836},
numpages = {13}
}

@misc{lu2025decentralizedretrievalaugmentedgeneration,
      title={A Decentralized Retrieval Augmented Generation System with Source Reliabilities Secured on Blockchain}, 
      author={Yining Lu and Wenyi Tang and Max Johnson and Taeho Jung and Meng Jiang},
      year={2025},
      eprint={2511.07577},
      archivePrefix={arXiv},
      primaryClass={cs.CR},
}

@misc{helmi2025decentralizingaimemoryshimi,
        title={Decentralizing AI Memory: SHIMI, a Semantic Hierarchical Memory Index for Scalable Agent Reasoning}, 
      author={Tooraj Helmi},
      year={2025},
      eprint={2504.06135},
      archivePrefix={arXiv},
      primaryClass={cs.AI},
}

@inproceedings{Proximity,
author = {Bergman, Shai and Kermarrec, Anne-Marie and Petrescu, Diana and Pires, Rafael and Randl, Mathis and de Vos, Martijn and Zhang, Ji},
title = {Leveraging Approximate Caching for Faster Retrieval-Augmented Generation},
year = {2025},
isbn = {9798400715549},
publisher = {Association for Computing Machinery},
address = {New York, NY, USA},
url = {https://doi.org/10.1145/3721462.3770776},
doi = {10.1145/3721462.3770776},
booktitle = {Proceedings of the 26th International Middleware Conference},
pages = {340–353},
numpages = {14},
location = {Vanderbilt University, Nashville, TN, USA},
series = {Middleware '25}
}

@inproceedings{6569858,
  author={Boutet, Antoine and Frey, Davide and Guerraoui, Rachid and Jégou, Arnaud and Kermarrec, Anne-Marie},
  booktitle={2013 IEEE 27th International Symposium on Parallel and Distributed Processing}, 
  title={WHATSUP: A Decentralized Instant News Recommender}, 
  year={2013},
  volume={},
  number={},
  pages={741-752},
  doi={10.1109/IPDPS.2013.47}
}

@inproceedings{9458649,
   author={Giakkoupis, George and Kermarrec, Anne-Marie and Ruas, Olivier and Taïani, François},
  booktitle={2021 IEEE 37th International Conference on Data Engineering (ICDE)}, 
  title={Cluster-and-Conquer: When Randomness Meets Graph Locality}, 
  year={2021},
  volume={},
  number={},
  pages={2027-2032},
  doi={10.1109/ICDE51399.2021.00195}
}

@article{Boutet2016,
author = {Boutet, Antoine and Frey, Davide and Guerraoui, Rachid and Jégou, Arnaud and Kermarrec, Anne-Marie},
year = {2015},
month = {03},
pages = {},
title = {Privacy-Preserving Distributed Collaborative Filtering},
volume = {8593},
isbn = {978-3-319-09580-6},
journal = {Computing},
doi = {10.1007/s00607-015-0451-z}
}

@article{FedVSE,
author = {Fan, Zeheng and Zeng, Yuxiang and Zheng, Zhuanglin and Tong, Yongxin},
title = {FedVSE: A Privacy-Preserving and Efficient Vector Search Engine for Federated Databases},
year = {2025},
issue_date = {August 2025},
publisher = {VLDB Endowment},
volume = {18},
number = {12},
issn = {2150-8097},
url = {https://doi.org/10.14778/3750601.3750674},
doi = {10.14778/3750601.3750674},
journal = {Proc. VLDB Endow.},
month = aug,
pages = {5371–5374},
numpages = {4}
}

@inproceedings{LSH-pstable-Dist,
author = {Datar, Mayur and Immorlica, Nicole and Indyk, Piotr and Mirrokni, Vahab S.},
title = {Locality-sensitive hashing scheme based on p-stable distributions},
year = {2004},
isbn = {1581138857},
publisher = {Association for Computing Machinery},
address = {New York, NY, USA},
url = {https://doi.org/10.1145/997817.997857},
doi = {10.1145/997817.997857},
booktitle = {Proceedings of the Twentieth Annual Symposium on Computational Geometry},
pages = {253–262},
numpages = {10},
location = {Brooklyn, New York, USA},
series = {SCG '04}
}

@misc{semantica,
      title={Semantica: Decentralized Search using a LLM-Guided Semantic Tree Overlay}, 
      author={Petru Neague and Quinten Stokkink and Naman Goel and Johan Pouwelse},
      year={2025},
      eprint={2502.10151},
      archivePrefix={arXiv},
      primaryClass={cs.IR},
}

@article{d-rag,
author = {Andersen, Tessa and Avalos, Ayanna and Dagher, Gaby and Long, Min},
year = {2025},
month = {02},
pages = {183-198},
title = {D-RAG: A Privacy-Preserving Framework for Decentralized RAG using Blockchain},
doi = {10.5121/csit.2025.150417}
}

@inproceedings{kademlia,
author = {Maymounkov, Petar and Mazi\`{e}res, David},
title = {Kademlia: A Peer-to-Peer Information System Based on the XOR Metric},
year = {2002},
isbn = {3540441794},
publisher = {Springer-Verlag},
address = {Berlin, Heidelberg},
booktitle = {Revised Papers from the First International Workshop on Peer-to-Peer Systems},
pages = {53–65},
numpages = {13},
series = {IPTPS '01}
}

@misc{drag/tarw,
      title={Distributed Retrieval-Augmented Generation}, 
      author={Chenhao Xu and Longxiang Gao and Yuan Miao and Xi Zheng},
      year={2025},
      eprint={2505.00443},
      archivePrefix={arXiv},
      primaryClass={cs.DC},
}

@misc{dragon,
      title={Efficient Distributed Retrieval-Augmented Generation for Enhancing Language Model Performance}, 
      author={Shangyu Liu and Zhenzhe Zheng and Xiaoyao Huang and Fan Wu and Guihai Chen and Jie Wu},
      year={2025},
      eprint={2504.11197},
      archivePrefix={arXiv},
      primaryClass={cs.LG},
}

@inproceedings{amazonDataset,
  author = {Ai, Qingyao and Zhang, Yongfeng and Bi, Keping and Chen, Xu and Croft, W. Bruce},
title = {Learning a Hierarchical Embedding Model for Personalized Product Search},
year = {2017},
isbn = {9781450350228},
publisher = {Association for Computing Machinery},
address = {New York, NY, USA},
url = {https://doi.org/10.1145/3077136.3080813},
doi = {10.1145/3077136.3080813},
booktitle = {Proceedings of the 40th International ACM SIGIR Conference on Research and Development in Information Retrieval},
pages = {645–654},
numpages = {10},
location = {Shinjuku, Tokyo, Japan},
series = {SIGIR '17}
}

@inproceedings{Pass2006Picture,
author = {Pass, Greg and Chowdhury, Abdur and Torgeson, Cayley},
title = {A picture of search},
year = {2006},
isbn = {1595934286},
publisher = {Association for Computing Machinery},
address = {New York, NY, USA},
url = {https://doi.org/10.1145/1146847.1146848},
doi = {10.1145/1146847.1146848},
booktitle = {Proceedings of the 1st International Conference on Scalable Information Systems},
pages = {1–es},
location = {Hong Kong},
series = {InfoScale '06}
}

@inproceedings{MacAvaney2022Reproducing,
  author = {MacAvaney, Sean and Macdonald, Craig and Ounis, Iadh},
title = {Reproducing Personalised Session Search Over the AOL Query Log},
year = {2022},
isbn = {978-3-030-99735-9},
publisher = {Springer-Verlag},
address = {Berlin, Heidelberg},
url = {https://doi.org/10.1007/978-3-030-99736-6_42},
doi = {10.1007/978-3-030-99736-6_42},
booktitle = {Advances in Information Retrieval: 44th European Conference on IR Research, ECIR 2022, Stavanger, Norway, April 10–14, 2022, Proceedings, Part I},
pages = {627–640},
numpages = {14},
location = {Stavanger, Norway}
}

@inproceedings{lewis2020retrieval,
author = {Lewis, Patrick and Perez, Ethan and Piktus, Aleksandra and Petroni, Fabio and Karpukhin, Vladimir and Goyal, Naman and K\"{u}ttler, Heinrich and Lewis, Mike and Yih, Wen-tau and Rockt\"{a}schel, Tim and Riedel, Sebastian and Kiela, Douwe},
title = {Retrieval-augmented generation for knowledge-intensive NLP tasks},
year = {2020},
isbn = {9781713829546},
publisher = {Curran Associates Inc.},
address = {Red Hook, NY, USA},
booktitle = {Proceedings of the 34th International Conference on Neural Information Processing Systems},
articleno = {793},
numpages = {16},
location = {Vancouver, BC, Canada},
series = {NeurIPS '20}
}

@inproceedings{reimers2019sentence,
author = {Reimers, Nils and Gurevych, Iryna},
year = {2019},
month = {01},
pages = {3973-3983},
title = {Sentence-BERT: Sentence Embeddings using Siamese BERT-Networks},
doi = {10.18653/v1/D19-1410}
}

@article{li2019approximate,
author={Li, Wen and Zhang, Ying and Sun, Yifang and Wang, Wei and Li, Mingjie and Zhang, Wenjie and Lin, Xuemin},
journal={ IEEE Transactions on Knowledge \& Data Engineering },
title={{ Approximate Nearest Neighbor Search on High Dimensional Data — Experiments, Analyses, and Improvement }},
year={2020},
volume={32},
number={08},
ISSN={1558-2191},
pages={1475-1488},
doi={10.1109/TKDE.2019.2909204},
url = {https://doi.ieeecomputersociety.org/10.1109/TKDE.2019.2909204},
publisher={IEEE Computer Society},
address={Los Alamitos, CA, USA},
month=aug
}

@article{ma2023comprehensive,
title={A Comprehensive Survey on Vector Database: Storage and Retrieval Technique, Challenge}, 
      author={Le Ma and Ran Zhang and Yikun Han and Shirui Yu and Zaitian Wang and Zhiyuan Ning and Jinghan Zhang and Ping Xu and Pengjiang Li and Ziyue Qiao and Wei Ju and Chong Chen and Dongjie Wang and Kunpeng Liu and Pengyang Wang and Pengfei Wang and Yanjie Fu and Chunjiang Liu and Yuanchun Zhou and Chang-Tien Lu},
      year={2026},
      eprint={2310.11703},
      archivePrefix={arXiv},
      primaryClass={cs.DB},
}

@inproceedings{tang2003peer,
author = {Tang, Chunqiang and Xu, Zhichen and Dwarkadas, Sandhya},
title = {Peer-to-peer information retrieval using self-organizing semantic overlay networks},
year = {2003},
isbn = {1581137354},
publisher = {Association for Computing Machinery},
address = {New York, NY, USA},
url = {https://doi.org/10.1145/863955.863976},
doi = {10.1145/863955.863976},
booktitle = {Proceedings of the 2003 Conference on Applications, Technologies, Architectures, and Protocols for Computer Communications},
pages = {175–186},
numpages = {12},
location = {Karlsruhe, Germany},
series = {SIGCOMM '03}
}

@article{maini2024llm,
      title={LLM Dataset Inference: Did you train on my dataset?}, 
      author={Pratyush Maini and Hengrui Jia and Nicolas Papernot and Adam Dziedzic},
      year={2024},
      eprint={2406.06443},
      archivePrefix={arXiv},
      primaryClass={cs.LG},
}

@article{rosati2025infringing,
title={Infringing AI: Liability for AI-Generated Outputs under International, EU, and UK Copyright Law}, volume={16}, DOI={10.1017/err.2024.72}, number={2}, journal={European Journal of Risk Regulation}, author={Rosati, Eleonora}, year={2025}, pages={603–627}
}

@article{amjad2026ai,
author = {Amjad, Isha},
year = {2026},
month = {02},
pages = {24-28},
title = {AI Copyright Infringement: Navigating the Legal Risks of AI-Generated Content},
volume = {4},
journal = {Trends in Intellectual Property Research},
doi = {10.69971/tipr.4.1.2026.104}
}

@inproceedings{zhuang2023toolqa,
author = {Zhuang, Yuchen and Yu, Yue and Wang, Kuan and Sun, Haotian and Zhang, Chao},
title = {ToolQA: a dataset for LLM question answering with external tools},
year = {2023},
publisher = {Curran Associates Inc.},
address = {Red Hook, NY, USA},
booktitle = {Proceedings of the 37th International Conference on Neural Information Processing Systems},
articleno = {2180},
numpages = {27},
location = {New Orleans, LA, USA},
series = {NeurIPS '23}
}

@inproceedings{brown2020language,
author = {Brown, Tom B. and Mann, Benjamin and Ryder, Nick and Subbiah, Melanie and Kaplan, Jared and Dhariwal, Prafulla and Neelakantan, Arvind and Shyam, Pranav and Sastry, Girish and Askell, Amanda and Agarwal, Sandhini and Herbert-Voss, Ariel and Krueger, Gretchen and Henighan, Tom and Child, Rewon and Ramesh, Aditya and Ziegler, Daniel M. and Wu, Jeffrey and Winter, Clemens and Hesse, Christopher and Chen, Mark and Sigler, Eric and Litwin, Mateusz and Gray, Scott and Chess, Benjamin and Clark, Jack and Berner, Christopher and McCandlish, Sam and Radford, Alec and Sutskever, Ilya and Amodei, Dario},
title = {Language models are few-shot learners},
year = {2020},
isbn = {9781713829546},
publisher = {Curran Associates Inc.},
address = {Red Hook, NY, USA},
booktitle = {Proceedings of the 34th International Conference on Neural Information Processing Systems},
articleno = {159},
numpages = {25},
location = {Vancouver, BC, Canada},
series = {NeurIPS '20}
}

@article{vaswani2017attention,
  title={Attention is all you need},
  author={Vaswani, Ashish and Shazeer, Noam and Parmar, Niki and Uszkoreit, Jakob and Jones, Llion and Gomez, Aidan N and Kaiser, {\L}ukasz and Polosukhin, Illia},
  journal={Advances in neural information processing systems},
  volume={30},
  year={2017}
}

@misc{alansari2026large,
  title={Large language models hallucination: A comprehensive survey},
  author={Alansari, Aisha and Luqman, Hamzah},
  eprint={2510.06265},
  archivePrefix={arXiv},
  primaryClass={cs.CL},
  year={2025}
}

@inproceedings{ji2023towards,
  title={Towards mitigating LLM hallucination via self reflection},
  author={Ji, Ziwei and Yu, Tiezheng and Xu, Yan and Lee, Nayeon and Ishii, Etsuko and Fung, Pascale},
  booktitle={Findings of the Association for Computational Linguistics: EMNLP 2023},
  pages={1827--1843},
  year={2023}
}

@inproceedings{lee2024one,
    title={One vs. many: Comprehending accurate information from multiple erroneous and inconsistent ai generations},
  author={Lee, Yoonjoo and Son, Kihoon and Kim, Tae Soo and Kim, Jisu and Chung, John Joon Young and Adar, Eytan and Kim, Juho},
  booktitle={Proceedings of the 2024 ACM Conference on Fairness, Accountability, and Transparency},
  pages={2518--2531},
  year={2024}
}

@article{nene1997simple,
  title={A simple algorithm for nearest neighbor search in high dimensions},
  author={Nene, Sameer A and Nayar, Shree K},
  journal={IEEE Transactions on pattern analysis and machine intelligence},
  volume={19},
  number={9},
  pages={989--1003},
  year={1997},
  publisher={IEEE}
}

@inproceedings{fan2024survey,
  title={A survey on rag meeting llms: Towards retrieval-augmented large language models},
  author={Fan, Wenqi and Ding, Yujuan and Ning, Liangbo and Wang, Shijie and Li, Hengyun and Yin, Dawei and Chua, Tat-Seng and Li, Qing},
  booktitle={Proceedings of the 30th ACM SIGKDD conference on knowledge discovery and data mining},
  pages={6491--6501},
  year={2024}
}

@article{zhao2026retrieval,
  title={Retrieval-augmented generation for ai-generated content: A survey},
  author={Zhao, Penghao and Zhang, Hailin and Yu, Qinhan and Wang, Zhengren and Geng, Yunteng and Fu, Fangcheng and Yang, Ling and Zhang, Wentao and Jiang, Jie and Cui, Bin},
  journal={Data Science and Engineering},
  pages={1--29},
  year={2026},
  publisher={Springer}
}

@article{rieke2020future,
   title={The future of digital health with federated learning},
  author={Rieke, Nicola and Hancox, Jonny and Li, Wenqi and Milletari, Fausto and Roth, Holger R and Albarqouni, Shadi and Bakas, Spyridon and Galtier, Mathieu N and Landman, Bennett A and Maier-Hein, Klaus and others},
  journal={NPJ digital medicine},
  volume={3},
  number={1},
  pages={119},
  year={2020},
  publisher={Nature Publishing Group UK London}
}

@misc{kuo2025research,
  title={Research in Collaborative Learning Does Not Serve Cross-Silo Federated Learning in Practice},
  author={Kuo, Kevin and Yadav, Chhavi and Smith, Virginia},
  eprint={2510.12595},
  archivePrefix={arXiv},
  primaryClass={cs.LG},
  year={2025}
}

@misc{hipaa1996,
  title        = {Health Insurance Portability and Accountability Act of 1996},
  author       = {{U.S. Department of Health \& Human Services}},
  year         = {1996},
  note         = {Public Law 104-191, 110 Stat. 1936},
  url          = {https://www.hhs.gov/hipaa}
}

@article{gilbert2002brewer,
  title={Brewer's conjecture and the feasibility of consistent, available, partition-tolerant web services},
  author={Gilbert, Seth and Lynch, Nancy},
  journal={Acm Sigact News},
  volume={33},
  number={2},
  pages={51--59},
  year={2002},
  publisher={ACM New York, NY, USA}
}

@inproceedings{jin2019pubmedqa,
  title={Pubmedqa: A dataset for biomedical research question answering},
  author={Jin, Qiao and Dhingra, Bhuwan and Liu, Zhengping and Cohen, William and Lu, Xinghua},
  booktitle={Proceedings of the 2019 conference on empirical methods in natural language processing and the 9th international joint conference on natural language processing (EMNLP-IJCNLP)},
  pages={2567--2577},
  year={2019}
}

@techreport{risson2007survey,
  title={Survey of research towards robust peer-to-peer networks: Search methods},
  author={Risson, John and Moors, Tim},
  year={2007}
}

@inproceedings{voulgaris2004exploiting,
  title={Exploiting semantic proximity in peer-to-peer content searching},
  author={Voulgaris, Spyros and Kermarrec, A-M and Massouli{\'e}, Laurent},
  booktitle={Proceedings. 10th IEEE International Workshop on Future Trends of Distributed Computing Systems, 2004. FTDCS 2004.},
  pages={238--243},
  year={2004},
  organization={IEEE}
}

@article{daniel2022ipfs,
  title={IPFS and friends: A qualitative comparison of next generation peer-to-peer data networks},
  author={Daniel, Erik and Tschorsch, Florian},
  journal={IEEE Communications Surveys \& Tutorials},
  volume={24},
  number={1},
  pages={31--52},
  year={2022},
  publisher={IEEE}
}

@inproceedings{guerraoui2024peerswap,
  title={Peerswap: A peer-sampler with randomness guarantees},
  author={Guerraoui, Rachid and Kermarrec, Anne-Marie and Kucherenko, Anastasiia and Pinot, Rafael and de Vos, Martijn},
  booktitle={2024 43rd International Symposium on Reliable Distributed Systems (SRDS)},
  pages={271--281},
  year={2024},
  organization={IEEE}
}

@article{jelasity2007gossip,
  title={Gossip-based peer sampling},
  author={Jelasity, M{\'a}rk and Voulgaris, Spyros and Guerraoui, Rachid and Kermarrec, Anne-Marie and Van Steen, Maarten},
  journal={ACM Transactions on Computer Systems (TOCS)},
  volume={25},
  number={3},
  pages={8--es},
  year={2007},
  publisher={ACM New York, NY, USA}
}

@article{STOKKINK2023109701,
  title={Web3 Sybil avoidance using network latency},
  author={Stokkink, Quinten and Ileri, Can Umut and Epema, Dick and Pouwelse, Johan},
  journal={Computer Networks},
  volume={227},
  pages={109701},
  year={2023},
  publisher={Elsevier}
}

@inproceedings{loser2004taxonomy,
  title={Taxonomy-based routing overlays in p2p networks},
  author={Loser, Alexander and Schubert, Kai and Zimmer, Frederik},
  booktitle={Proceedings. International Database Engineering and Applications Symposium, 2004. IDEAS'04.},
  pages={407--412},
  year={2004},
  organization={IEEE}
}

@inproceedings{shen2006building,
  title={Building bridges for web query classification},
  author={Shen, Dou and Sun, Jian-Tao and Yang, Qiang and Chen, Zheng},
  booktitle={Proceedings of the 29th annual international ACM SIGIR conference on Research and development in information retrieval},
  pages={131--138},
  year={2006}
}

@misc{allmpnetbasev2,
  author       = "{Sentence Transformers}",
  title        = "all-mpnet-base-v2",
  howpublished = {\url{https://huggingface.co/sentence-transformers/all-mpnet-base-v2}},
}

@article{abdin2024phi,
      title={Phi-3 Technical Report: A Highly Capable Language Model Locally on Your Phone}, 
      author={Marah Abdin and Jyoti Aneja and Hany Awadalla and Ahmed Awadallah and Ammar Ahmad Awan and Nguyen Bach and Amit Bahree and Arash Bakhtiari and Jianmin Bao and Harkirat Behl and Alon Benhaim and Misha Bilenko and Johan Bjorck and Sébastien Bubeck and Martin Cai and Qin Cai and Vishrav Chaudhary and Dong Chen and Dongdong Chen and Weizhu Chen and Yen-Chun Chen and Yi-Ling Chen and Hao Cheng and Parul Chopra and Xiyang Dai and Matthew Dixon and Ronen Eldan and Victor Fragoso and Jianfeng Gao and Mei Gao and Min Gao and Amit Garg and Allie Del Giorno and Abhishek Goswami and Suriya Gunasekar and Emman Haider and Junheng Hao and Russell J. Hewett and Wenxiang Hu and Jamie Huynh and Dan Iter and Sam Ade Jacobs and Mojan Javaheripi and Xin Jin and Nikos Karampatziakis and Piero Kauffmann and Mahoud Khademi and Dongwoo Kim and Young Jin Kim and Lev Kurilenko and James R. Lee and Yin Tat Lee and Yuanzhi Li and Yunsheng Li and Chen Liang and Lars Liden and Xihui Lin and Zeqi Lin and Ce Liu and Liyuan Liu and Mengchen Liu and Weishung Liu and Xiaodong Liu and Chong Luo and Piyush Madan and Ali Mahmoudzadeh and David Majercak and Matt Mazzola and Caio César Teodoro Mendes and Arindam Mitra and Hardik Modi and Anh Nguyen and Brandon Norick and Barun Patra and Daniel Perez-Becker and Thomas Portet and Reid Pryzant and Heyang Qin and Marko Radmilac and Liliang Ren and Gustavo de Rosa and Corby Rosset and Sambudha Roy and Olatunji Ruwase and Olli Saarikivi and Amin Saied and Adil Salim and Michael Santacroce and Shital Shah and Ning Shang and Hiteshi Sharma and Yelong Shen and Swadheen Shukla and Xia Song and Masahiro Tanaka and Andrea Tupini and Praneetha Vaddamanu and Chunyu Wang and Guanhua Wang and Lijuan Wang and Shuohang Wang and Xin Wang and Yu Wang and Rachel Ward and Wen Wen and Philipp Witte and Haiping Wu and Xiaoxia Wu and Michael Wyatt and Bin Xiao and Can Xu and Jiahang Xu and Weijian Xu and Jilong Xue and Sonali Yadav and Fan Yang and Jianwei Yang and Yifan Yang and Ziyi Yang and Donghan Yu and Lu Yuan and Chenruidong Zhang and Cyril Zhang and Jianwen Zhang and Li Lyna Zhang and Yi Zhang and Yue Zhang and Yunan Zhang and Xiren Zhou},
      year={2024},
      eprint={2404.14219},
      archivePrefix={arXiv},
      primaryClass={cs.CL},
}

@article{nakamoto2008bitcoin,
  title={Bitcoin: A peer-to-peer electronic cash system},
  author={Nakamoto, Satoshi},
  year={2008}
}

@inproceedings{pouwelse2005bittorrent,
author = {Pouwelse, J.A. and Garbacki, Pawel and Epema, D. and Sips, Henk},
year = {2005},
month = {02},
pages = {205-216},
title = {The Bittorrent P2P File-Sharing System: Measurements and Analysis},
volume = {3640},
isbn = {978-3-540-29068-1},
journal = {Peer-to-Peer Systems IV},
doi = {10.1007/11558989_19}
}

@INPROCEEDINGS{sripanidkulchai2003efficient,
  author={Sripanidkulchai, K. and Maggs, B. and Zhang, H.},
  booktitle={IEEE INFOCOM 2003. Twenty-second Annual Joint Conference of the IEEE Computer and Communications Societies (IEEE Cat. No.03CH37428)}, 
  title={Efficient content location using interest-based locality in peer-to-peer systems}, 
  year={2003},
  volume={3},
  number={},
  pages={2166-2176 vol.3},
  doi={10.1109/INFCOM.2003.1209237}}

@inproceedings{loeser2005semantic,
	address = {Berlin, Heidelberg},
	author = {L{\"o}ser, Alexander and Staab, Steffen and Tempich, Christoph},
	booktitle = {Multiagent System Technologies},
	editor = {Eymann, Torsten and Kl{\"u}gl, Franziska and Lamersdorf, Winfried and Klusch, Matthias and Huhns, Michael N.},
	isbn = {978-3-540-28741-4},
	pages = {15--26},
	publisher = {Springer Berlin Heidelberg},
	title = {Semantic Methods for P2P Query Routing},
	year = {2005}}

@article{abdelhamid2023lightweight,
  title={A lightweight anomaly detection system for black hole attack},
  author={Abdelhamid, Ashraf and Elsayed, Mahmoud Said and Jurcut, Anca D and Azer, Marianne A},
  journal={Electronics},
  volume={12},
  number={6},
  pages={1294},
  year={2023},
  publisher={MDPI}
}

@inproceedings{douceur2002sybil,
  title={The sybil attack},
  author={Douceur, John R},
  booktitle={International workshop on peer-to-peer systems},
  pages={251--260},
  year={2002},
  organization={Springer}
}

@inproceedings{heilman2015eclipse,
  title={Eclipse attacks on $\{$Bitcoin’s$\}$$\{$peer-to-peer$\}$ network},
  author={Heilman, Ethan and Kendler, Alison and Zohar, Aviv and Goldberg, Sharon},
  booktitle={24th USENIX security symposium (USENIX security 15)},
  pages={129--144},
  year={2015}
}

\end{document}